\documentclass[aps,prd,preprint,superscriptaddress,showpacs]{revtex4}
\usepackage{amsfonts,amssymb,amsmath}
\usepackage{graphicx}

\newcommand{\be}{\begin{equation}}
\newcommand{\bea}{\begin{eqnarray}}
\newcommand{\ee}{\end{equation}}
\newcommand{\eea}{\end{eqnarray}}
\newcommand{\bpi}{\begin{picture}}
\newcommand{\bce}{\begin{center}}
\newcommand{\epi}{\end{picture}}
\newcommand{\ece}{\end{center}}
\newcommand{\D}{\displaystyle}
\def\chic#1{{\scriptscriptstyle #1}}

\def\g{\widetilde{{\rm I}\hspace{-0.07cm}\Gamma}}
\def\gv{\widetilde{{\rm I}\hspace{-0.07cm}\Gamma}}

\def\gnp{{g}^2}
\renewcommand{\theequation}{\arabic{section}.\arabic{equation}}
\newcommand{\de}{{\Delta}_{\rm \mbox{\tiny E}}}  
\newcommand{\da}{\rm \mbox{\tiny E}}  

\newcommand{\Valencia}{Departamento de F\'\i sica Te\'orica
and IFIC, Centro Mixto,\\ Universidad de Valencia -- CSIC \\
E-46100, Burjassot, Valencia, Spain}

\begin{document}


\title{Power-law running of the effective gluon mass}

\author{Arlene~C.~Aguilar}
\affiliation{\Valencia}

\author{Joannis Papavassiliou}
\affiliation{\Valencia}

\begin{abstract}

The  dynamically generated  effective gluon  mass is  known  to depend
non-trivially  on the  momentum, decreasing  sufficiently fast  in the
deep  ultraviolet, in  order for  the renormalizability  of QCD  to be
preserved. General arguments based on the analogy with the constituent
quark   masses,   as  well   as   explicit   calculations  using   the
operator-product expansion,  suggest that the gluon mass  falls off as
the inverse square of the momentum, relating it to the gauge-invariant
gluon condensate  of dimension four.   In this article  we demonstrate
that  the power-law  running of  the  effective gluon  mass is  indeed
dynamically   realized   at   the   level  of   the   non-perturbative
Schwinger-Dyson  equation.   We  study  a  gauge-invariant  non-linear
integral equation  involving the gluon self-energy,  and establish the
conditions necessary  for the existence of  infrared finite solutions,
described  in terms of  a momentum-dependent  gluon mass.   Assuming a
simplified  form  for the  gluon  propagator,  we  derive a  secondary
integral equation  that controls the running  of the mass  in the deep
ultraviolet.  Depending  on the  values chosen for  certain parameters
entering  into the  Ansatz for  the fully-dressed  three-gluon vertex,
this  latter equation  yields either  logarithmic  solutions, familiar
from previous  linear studies, or  a new type of  solutions, displaying
power-law running.   In addition, it furnishes  a non-trivial integral
constraint,  which  restricts significantly  (but  does not  determine
fully)  the running  of  the  mass in  the  intermediate and  infrared
regimes.  The  numerical analysis  presented is in  complete agreement
with the analytic results  obtained, showing clearly the appearance of
the two  types of momentum-dependence, well-separated  in the relevant
space  of parameters.   Several technical  improvements,  various open
issues, and possible future directions, are briefly discussed.

\end{abstract}


\maketitle

\setcounter{section}{0}
\section{Introduction}
\label{Sect:Intro}

The possibility that the  non-perturbative dynamics of QCD generate an
effective  gluon mass,  first  elaborated in  the  pioneering work  of
Cornwall~\cite{Cornwall:1979hz,Cornwall:1981zr}, has received considerable 
theoretical~\cite{Bernard:1982my,Kondo:2001nq,
Bloch:2003yu,Aguilar:2004sw,Dudal:2004rx,Aguilar:2006gr}
and  phenomenological~\cite{Parisi:1980jy,Mattingly:1992ud,Mihara:2000wf}
attention in  recent  years~\cite{lattice}.   
As advocated  in
\cite{Cornwall:1981zr}, the  origin of  this effective mass  is purely
dynamical   and   preserves   the   local  $SU(3)_c$   invariance   of
QCD,  in   close  analogy  to  what   happens  in  QED$_2$
(Schwinger model)~\cite{Schwinger:1962tn}, 
where the  photon acquires a mass without violating
the  abelian  gauge  symmetry.   The  gluon mass  is  not  a  directly
measurable  quantity,  and  its  value  is  determined,  at  least  in
principle,  by  relating it  to  other dimensionful  non-perturbative
parameters,  such  as  the  string  tension,  glueball  masses,  gluon
condensates, and the vacuum energy of QCD~\cite{Shifman:1978by}.

Since gluon  mass generation is a purely  non-perturbative effect, the
most  standard way for  studying it  in the  continuum is  through the
Schwinger-Dyson  equations   (SDE)  governing  the   relevant  Green's
functions.  These equations have been treated from the
very beginning~\cite{Cornwall:1981zr} in a manifestly gauge-invariant 
way, by resorting to the systematic rearrangement 
of graphs implemented by the pinch technique  (PT)
~\cite{Cornwall:1981zr,Cornwall:1989gv,Binosi:2002ft}.
Subsequently, the powerful all-order  connection~\cite{Binosi:2002ft} 
between the PT and the Feynman gauge of the Background Field  Method (BFM)~\cite{Abbott:1980hw}
gave rise to the ``PT-BFM'' truncation scheme~\cite{Aguilar:2006gr,Binosi:2006da} 
that guarantees crucial properties,  such as gauge-invariance,
gauge-independence,  and  invariance  under the  renormalization-group (RG).  
What one characterizes as gluon mass generation at
the  level of  the  SDE governing the PT-BFM 
gluon self-energy  is
essentially   the  existence   of  solutions   that  reach   a  finite
(non-vanishing) value in the  deep infrared (IR).  These solutions may
be  successfully   fitted  by  a   ``massive''  propagator  of   the 
form $\Delta^{-1}(q^2)  =  q^2  +  m^2(q^2)$; the  crucial  characteristic,
enforced by  the SDE itself, is  that $m^2(q^2)$ is  not ``hard'', but
depends non-trivially  on the momentum  transfer $q^2$.  Specifically,
$m^2(q^2)$  is  a monotonically  decreasing  function,  starting at  a
non-zero   value  in   the  IR  ($m^2(0)   >0$)   and  dropping
``sufficiently  fast'' in  the  deep ultraviolet  (UV).   When the  RG
logarithms are  properly taken into  account, one obtains  in addition
the  non-perturbative  generalization  of  $g^2(q^2)$, 
the  QCD  running  coupling (effective charge).
The presence of $m^2(q^2)$  in the corresponding
logarithms   tames  the   Landau  singularity   associated   with  the
perturbative $\beta$  function, and the resulting  effective charge is
asymptotically free in  the UV, ``freezing'' at a  finite value in the
IR~\cite{Cornwall:1981zr,Cornwall:1989gv,Papavassiliou:1991hx,Badalian:1999fq,
Aguilar:2002tc,Brodsky:2002nb,Mattingly:1993ej}.

The  running   of  $m^2(q^2)$  is   of  central  importance   for  the
self-consistency of this entire approach.  Roughly speaking, the value
of   $\Delta^{-1}(0)$    is   determined   by    integrals   involving
$\Delta(q^2)$,  $m^2(q^2)$, and  $g^2(q^2)$ over  the entire  range of
(Euclidean)  momenta. The  UV convergence  of these  integrals depends
crucially on  how $m^2(q^2)$ behaves as  $q^2\to\infty$. If $m^2(q^2)$
drops   off    asymptotically   faster   than    a   logarithm,   then
$\Delta^{-1}(0)$ is finite. This  is crucial because the finiteness of
$\Delta^{-1}(0)$  guarantees   essentially  the  renormalizability  of
QCD. Had the mass been constant instead, these integrals would diverge
quadratically; to absorb such a divergence one would have to introduce
a counterterm of the form  $ m^2_0 (\Lambda^2_{{\chic U}{\chic V}}) A^2$
at  the level  of the  fundamental  QCD Lagrangian,  which is  clearly
forbidden by the local gauge invariance.

The UV behavior  of $m^2(q^2)$ is not imposed by  hand, but is instead
determined dynamically  from the corresponding SDE.   The situation is
conceptually  very similar to  the case  of the  dynamically generated
(constituent) quark masses, whose momentum dependence is controlled by
the  gap  equation.   Notice,  however,  that there  is  an  important
technical difference. Due to its  Dirac structure the gap equation may
be separated into two  independent components, one determining the wave
function and one the mass of the quark self-energy. In the case of the
SDE  for the  gluon self-energy  there is  no such  direct separation:
instead an appropriate matching of  the contributions on both sides of
the equations must be carried out.

In      previous     studies      of     linear      SDE     equations
~\cite{Cornwall:1981zr,Aguilar:2006gr}  the UV running  of $m^2(q^2)$
has been found  to be logarithmic, of the  general form $m^2(q^2) \sim
(\ln   q^2)^{-1-\gamma}$,  with   $\gamma   >0$  in   order  for   the
aforementioned integrals determining $\Delta^{-1}(0)$ to converge.  Of
course, it is natural to ask  whether the QCD dynamics allows also for
a  $m^2(q^2)$  displaying  power-law  running,  i.e.   $m^2(q^2)  \sim
q^{-2}(\ln q^2)^{\gamma-1}$, as happens in the case of the dynamically
generated quark masses~\cite{Lane:1974he}.  This possibility was first
envisaged by Cornwall~\cite{Cornwall:1981zr,Cornwall:1985bg}, based on
the  aforementioned studies  of  chiral symmetry  breaking.  A  decade
later,  Lavelle~\cite{Lavelle:1991ve} wrote down  the operator-product
expansion  (OPE)  for  the   (partially  dressed)  one-loop  PT  gluon
self-energy,  expressing it  solely  in terms  of the  gauge-invariant
gluon condensate $\langle  G^2 \rangle = \langle 0|\!:\!G_{\mu\nu}^{a}
G^{\mu\nu}_{a}\!:\!|0  \rangle$  of  dimension  four (no  quarks  were
considered)~\cite{com3}.    The   resulting   self-energy   was   then
identified  with an  effective gluon  mass, $m^2(q^2)\sim  \langle G^2
\rangle/q^2$, i.e.  a mass  displaying power-law running.  However, to
date, the power-law running of the effective gluon mass has never been
demonstrated as  an explicit dynamical  possibility at the level  of a
(non-linear) SDE.

In  this  article  we show  that  the  non-linear  SDE for  the  gluon
self-energy  in  the  PT-BFM  formalism  has  two  distinct  types  of
solutions: (i) solutions of the type already encountered in the linear
studies, with $m^2(q^2)$  running as an inverse power  of a logarithm,
and (ii) solutions found for the first time, where the effective gluon
mass drops  asymptotically as an  {\it inverse power of  the momentum}
(multiplied  by powers  of logarithms).   Which  of the  two types  of
solutions  will  be  actually  realized  is  a  complicated  dynamical
problem,  depending  mainly on  the  details  of  the (fully  dressed)
three-gluon vertex, $\gv_{\mu\alpha\beta}$,  entering into the SDE for
the gluon self-energy (see Fig.~\ref{fa}).

The article is organized as follows: 
In Sec.~\ref{Sect:full}, after setting up the notation,
 we present the derivation of the 
SDE for the PT-BFM gluon propagator. 
We restrict our 
analysis to the subset of ``one-loop dressed'' gluonic contributions,
namely the two fully-dressed gluonic diagrams 
shown in Fig.~\ref{fa}. In addition, we explain briefly how the  
naive all-order Ward identity (WI) satisfied by the 
full three-gluon vertex in this formalism enforces the transversality of the 
gluon self-energy even in the absence of ghost loops. 
In Sec.~\ref{Sect:IRF} we manipulate 
the SDE derived in the previous section further,
with the motivation to search  for infrared-finite solutions.
An important ingredient at this stage is the 
Ansatz introduced for the full three gluon vertex, based on 
the gauge-technique; this vertex satisfies, by construction, 
the simple all-order WI characteristic of the PT-BFM,   
and contains sufficient structure to give rise to a non-vanishing  $\Delta^{-1}(0)$.
In addition, we discuss in detail  
the technical adjustments implemented to the SDE, in order to 
endow it with the correct RG behavior. 
The next two sections contain the main results of this article.
Specifically, in Sec.~\ref{Sect:UVMASS} we extract from the SDE
an integral equation that determines the running 
of the effective gluon mass in the UV, and study its solutions.
We derive an important constraint, in the form of an integral boundary condition,
relating the value of $\Delta^{-1}(0)$ with 
$m^2(q^2)$ in the entire range of momenta; 
this condition 
must be necessarily satisfied in order for the mass equation to 
have solutions that vanish asymptotically in the deep UV.
Then we demonstrate that, depending on the values 
of two basic parameters, one finds solutions for the 
masses that drop 
as inverse powers of a logarithm of $q^2$, or much faster, as an inverse power 
of $q^2$. 
In Sec.~\ref{Sect:NUM} we carry out a numerical analysis of the SDE, which 
fully corroborates the existence of the two aforementioned types 
of solutions.  
In Sec.~\ref{Sect:Concl} 
we present our conclusions and a discussion of various open issues. 
Finally, in an Appendix
we present all technical points related to the modifications 
one must introduce to the standard angular approximation 
in order to capture correctly the leading $m^2(q^2)$ behavior,
encoded in the exact SDE.  


\setcounter{equation}{0}
\section{SDE for the PT-BFM gluon propagator}
\label{Sect:full}

\begin{figure}[ht]
\includegraphics[scale=0.8]{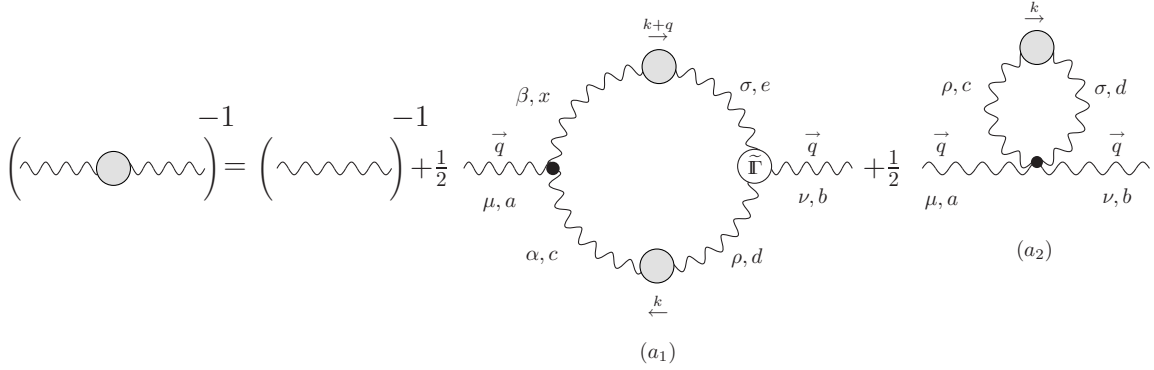}
\caption{The gluonic ``one-loop dressed'' contributions to the SDE.}
\label{fa}
\end{figure}

In this  section we derive in detail a  
non-linear SDE equation for the gluon
propagator in the PT-BFM formalism.
As  has  been explained  in  the
literature,  this  scheme  is  essentially founded  on  the  all-order
correspondence   between    PT   and   BFM~\cite{Binosi:2002ft}:   the
(gauge-independent) PT effective $n$-point functions coincide with the
(gauge-dependent) BFM $n$-point functions provided that the latter are
computed in the  Feynman gauge.  One of the  most powerful features of
this formalism is  the special way in which  the transversality of the
all-order   PT-BFM   self-energy  is   realized.
Specifically,  by virtue  of  the Abelian-like  WIs  satisfied by  the
vertices   involved,  gluonic   and  ghost   contributions   are  {\it
separately}   transverse,    within   {\it   each}    order   in   the
``dressed-loop'' expansion of the corresponding SDE~\cite{Aguilar:2006gr}.
This property, in turn, allows for a systematic truncation of the full
SDE~\cite{Sohn:1985em} that preserves  the  crucial property  of  gauge invariance.   In
particular,  instead of a  system of  two coupled  equations involving
gluon  and  ghost  propagators,  one  may  consider  only  the  subset
containing  gluons,  without compromising  the  transversality of  the
gluon self-energy.   Therefore, in what follows,  
we  will consider only
the gauge-invariant subset of {\it ``one-loop dressed'' gluonic} diagrams,
given by the two graphs of  Fig.~\ref{fa}.

Within this formalism  there are two distinct gluon propagators,
the  background gluon propagator
$\widehat{\Delta}_{\mu\nu}(q)$ ( which, in the Feynman gauge, coincides  
with the PT gluon propagator) and 
the quantum gluon propagator ${\Delta}_{\mu\nu}(q)$,
appearing inside the loops.
By virtue of a powerful all-order identity~\cite{Gambino:1999ai,Binosi:2002ez}, 
one may express ${\Delta}_{\mu\nu}(q)$ in terms
of $\widehat{\Delta}_{\mu\nu}(q)$ and auxiliary (unphysical) Green's functions
involving anti-fields and background sources.
As a first approximation, in this work we will neglect
the effects of the aforementioned auxiliary Green's functions, 
and carry out the substitution ${\Delta}_{\mu\nu}(q) \to \widehat{\Delta}_{\mu\nu}(q)$ throughout
(and drop the ``hats'' to simplify the notation).
 
In the Feynman gauge (both of the usual linear renormalizable gauges as well as the BFM)
the full gluon propagator $\Delta_{\mu\nu}(q)$ has the general form (we suppress color indices)
\begin{equation}
\Delta_{\mu\nu}(q)= {-\D i}\left[{\rm P}_{\mu\nu}(q)\Delta(q^2) + 
\frac{q_{\mu}q_{\nu}}{q^4}\right]\,,
\label{prop_cov}
\end{equation}
where 
\be
{\rm P}_{\mu\nu}(q)= \ g_{\mu\nu} - \frac{\D q_\mu q_\nu}{\D q^2}\,,
\label{projector}
\ee
is the usual transverse projector.
The scalar function $\Delta(q^2)$ is related to the 
all-order gluon self-energy $\Pi_{\mu\nu}(q)$,
\be
\Pi_{\mu\nu}(q)={\rm P}_{\mu\nu}(q) \Pi(q^2)\,,
\ee
through
\be
\Delta^{-1}(q^2) = q^2 + i \Pi(q^2)\,.
\label{fprog}
\ee
Notice that, since 
$\Pi(q^2)$ has been defined in (\ref{fprog}) 
with the imaginary  factor $i$ pulled out in front, it is   
given simply by the corresponding Feynman diagrams in Minkowski space.
Finally, the inverse of the full gluon propagator is given by
\bea
\Delta^{-1}_{\mu\nu}(q) &=& ig_{\mu\nu}q^2 - \Pi_{\mu\nu}(q)
\nonumber\\
&=& i{\rm P}_{\mu\nu}(q) \Delta^{-1}(q^2) + iq_{\mu}q_{\nu}\,.
\label{inv_prog}
\eea
At tree-level,
\be
\Delta_{0}^{\mu\nu}(q) = \frac{g^{\mu\nu}}{q^2} \,,
\,\,\,\,\,\,
\Delta_{0}(q^2) = \frac{1}{q^2}\,.
\ee

The SDE in Fig.~\ref{fa} reads,
\be
({\Delta}^{-1})_{\mu\nu}^{ab}(q)
= iq^2 g_{\mu\nu} \delta^{ab} -  
\left[{\Pi}^{ab}_{\mu\nu}(q)
\big|_{{a_1}} + {\Pi}^{ab}_{\mu\nu}
\big|_{{a_2}}
\right]\,,
\label{SD1}
\ee
with
the closed expressions corresponding to the diagrams $(a_1)$ and $(a_2)$ given by 
\begin{eqnarray}
{\Pi}^{ab}_{\mu\nu}(q) \big|_{{a_1}} &=& 
\frac{1}{2} \, \int\![dk]\,
\widetilde{\Gamma}_{\mu\alpha\beta}^{acx}(q,k,-k-q)
\Delta^{\alpha\rho}_{cd}(k)
{\g}_{\nu\rho\sigma}^{bde}(-q,-k,k+q)
\Delta^{\beta\sigma}_{xe}(k+q)\,,
\nonumber\\
{\Pi}^{ab}_{\mu\nu}\big|_{{a_2}} &=&
\frac{1}{2} \,\int\! [dk]\,
\widetilde{\Gamma}_{\mu\nu\rho\sigma}^{abcd}
\Delta^{\rho\sigma}_{cd} (k)\,.
\label{sde0}
\end{eqnarray}
The flow of momenta, together with the color and Lorentz indices, are shown  
in Fig.~\ref{fa}. 
We have assumed dimensional regularization, employing the short-hand notation
\mbox{$ [dk] =  d^d k/(2\pi)^d$}, 
where $d=4-\epsilon$ is the dimension of space-time.

We will use greek letters for the Lorentz indices and latin for the color indices.
The tree-level vertex $\widetilde{\Gamma}_{\mu\alpha\beta}^{acx}$ appearing in $(a_1)$
is given by ($p_1=k$, $p_2=-k-q$)
\bea
\widetilde{\Gamma}_{\mu\alpha\beta}^{acx}(q,p_1,p_2)&=& 
g f^{acx}\widetilde{\Gamma}_{\mu\alpha\beta}(q,p_1,p_2), \nonumber\\ 
\widetilde{\Gamma}_{\mu\alpha\beta}(q,p_1,p_2)&=&  
(p_1-p_2)_{\mu} g_{\alpha\beta} + 2q_{\beta}g_{\mu\alpha} - 2q_{\alpha}g_{\mu\beta},
\label{gfey}
\eea
and satisfies the elementary WI
\be
q^{\mu}\widetilde{\Gamma}_{\mu\alpha\beta}(q,p_1,p_2)=
(p_2^2 - p_1^2) g_{\alpha\beta} = 
i\left[\Delta_{0\,\alpha\beta}^{-1}(p_1)- 
\Delta_{0\,\alpha\beta}^{-1}(p_2)\right]
= \left[\Delta_{0}^{-1}(p_2)- 
\Delta_{0}^{-1}(p_1)\right] g_{\alpha\beta}\,.
\label{tree_ward}
\ee
The bare four-gluon vertex $\widetilde{\Gamma}_{\mu\nu\rho\sigma}^{abcd}$ appearing in $(a_2)$ 
is given by 
\bea
\widetilde{\Gamma}_{\mu\nu\rho\sigma}^{abcd}
&=&-ig^2\bigg[
f^{acx} f^{xbd}\left(g_{\mu\nu}g_{\rho\sigma}-g_{\mu\sigma}g_{\rho\nu}
+ g_{\mu\rho}g_{\nu\sigma}\right) +  f^{adx} f^{xcb}\left( g_{\mu\rho}g_{\nu\sigma}-g_{\mu\nu}g_{\rho\sigma}
- g_{\mu\sigma}g_{\rho\nu} \right) \bigg.  \nonumber \\
 &&+ \bigg. f^{abx} f^{xcd}
\left(g_{\mu\rho}g_{\nu\sigma}-g_{\mu\sigma}g_{\rho\nu}\right) \bigg]\,.
\label{bfm4gluon}
\eea
In addition, for the fully dressed quantities ${\Delta}_{\mu\nu}^{ab}$ and  $\gv_{\mu\alpha\beta}^{abc}$
we will set
${\Delta}_{\mu\nu}^{ab}= \delta^{ab} {\Delta}_{\mu\nu}$
and $\gv_{\mu\alpha\beta}^{abc}= gf^{abc}\gv_{\mu\alpha\beta}$. 
In the  PT-BFM formalism the all-order WI satisfied 
by $\gv_{\mu\alpha\beta}$ is the naive generalization of 
the tree-level WI (\ref{tree_ward})~\cite{Abbott:1980hw,Aguilar:2006gr}, i.e.
\begin{equation}
q^{\mu}\gv_{\mu\alpha\beta}(q,p_1,p_2) = 
i[{\Delta}^{-1}_{\alpha\beta}(p_1) - {\Delta}^{-1}_{\alpha\beta}(p_2)]\,.
\label{ward1}
\end{equation}
It is then elementary to verify that 
\be
q^{\nu}\left[ {\Pi}_{\mu\nu}(q)\big|_{a_1}+ {\Pi}_{\mu\nu}\big|_{a_2}\right] = 0\,.
\ee
Thus, 
the subset of graphs considered is transverse by itself, 
i.e. without the inclusion of ghost loops, as announced.
 
In order to reduce the algebraic complexity 
of the problem, we drop the longitudinal terms  
from the gluon propagators inside the integrals  on the  r.h.s. of (\ref{sde0}), i.e. we set
${\Delta}_{\alpha\beta} \to -ig_{\alpha\beta} {\Delta}$.
This may be done 
without compromising the transversality of the answer, provided that one drops, 
at the same time, the 
longitudinal pieces on the  r.h.s. of the WI of Eq.(\ref{ward1})~\cite{foot1}.

It is then straightforward to arrive at the following form for the SDE
\be
i {\rm P}_{\mu\nu}(q) \Delta^{-1}(q^2)  
=i {\rm P}_{\mu\nu}(q) \,q^2 - \frac{C_{\rm A} g^2}{2}\,
 \Bigg( \int\!  [dk]\,
\widetilde{\Gamma}_{\mu}^{\alpha\beta}
{\Delta}(k)
{\g}_{\nu\alpha\beta} 
{\Delta}(k+q)
-\,8 \, g_{\mu\nu}
\int\!  [dk] \, {\Delta}(k) \Bigg)\,,
\label{polar2}
\ee
where we have used that
$f^{ace}f^{bce}= \delta^{ab} C_{\rm A}$,
with $C_{\rm A}$ the  Casimir eigenvalue in the adjoint representation [$C_{\rm A}=N$ for $SU(N)$].
After the  omission of the longitudinal parts
the WI of (\ref{ward1}) becomes
\be
q^{\nu} {\g}_{\nu\alpha\beta}(q,p_1,p_2) = 
 \left[{\Delta}^{-1}(p_2) - {\Delta}^{-1}(p_1)\right] g_{\alpha\beta}\,;
\label{WID}
\ee
contracting by $q^{\nu}$ and using (\ref{WID}) 
one may easily  verify 
that the sum of the two integrals on the r.h.s. of (\ref{polar2}) is indeed transverse.

\setcounter{equation}{0}
\section{SDE with IR-finite solutions}
\label{Sect:IRF}

In order to proceed further with Eq.(\ref{polar2}), one needs to supply some information 
about the form of the full vertex $\gv$. To accomplish this,
ideally one should set up the corresponding SDE governing the 
vertex $\gv$, and solve a system of coupled integral equations.
In practice this is very difficult and one 
almost always resorts to the ``gauge technique"~\cite{Salam:1963sa}, expressing 
$\gv^{\mu\alpha\beta}$ as a functional of $\Delta$, 
in such a way as to satisfy (by construction) the appropriate WI.   
It is clear that this procedure leaves the transverse 
(i.e. identically conserved) part of the vertex undetermined, a fact that leads 
to the mishandling of overlapping divergences, and forces one 
to renormalize the resulting SD equation subtractively instead of multiplicatively.
 
The Ansatz  we will use for the vertex is 

\be
\gv^{\mu\alpha\beta}= 
L^{\mu\alpha\beta} + T_1^{\mu\alpha\beta} + T_2^{\mu\alpha\beta}\,,
\label{gtvertex}
\ee
with 
\bea
L^{\mu\alpha\beta}(q,p_1,p_2)  &=&  \widetilde{\Gamma}^{\mu\alpha\beta}(q,p_1,p_2)
+i g^{\alpha\beta}\, \frac{q^{\mu}}{q^2}\,
\left[ {\Pi}(p_2) - {\Pi}(p_1)\right]\,,
\nonumber\\
T_1^{\mu\alpha\beta}(q,p_1,p_2) &=& 
- i\frac{c_1}{q^2}\left(q^{\beta}g^{\mu\alpha} - q^{\alpha}g^{\mu\beta}\right)
\left[{\Pi}(p_1)+ {\Pi}(p_2)\right]\,,\nonumber\\
T_2^{\mu\alpha\beta}(q,p_1,p_2) &=& -i c_2 \left(q^{\beta}g^{\mu\alpha} - q^{\alpha}g^{\mu\beta}\right)
\left[\frac{{\Pi}(p_1)}{p_1^2}+ \frac{{\Pi}(p_2)}{p_2^2}\right]\,.
\label{LT1T2}
\eea

Several comments on the properties and role of the vertex $\gv^{\mu\alpha\beta}$ and its 
individual components are now in order:

({\bf i})~ When ${\Pi}(p_i)=0$,  $\gv^{\mu\alpha\beta}$ 
goes over to the tree-level (bare) result, namely  
the $\widetilde{\Gamma}^{\mu\alpha\beta}$ of (\ref{gfey}).

({\bf ii})~ $\gv^{\mu\alpha\beta}$ 
is Bose symmetric only with respect to its two legs appearing inside the loop, 
carrying momentum $p_1=k$ and  $p_2=-(k+q)$; so, 
$\gv^{\mu\alpha\beta}(q,p_1,p_2)$ is invariant under the 
simultaneous
exchange  $p_1 \longleftrightarrow p_2$,  $\alpha \longleftrightarrow\beta$ and 
$b\longleftrightarrow c$.

({\bf iii})~ $\gv^{\mu\alpha\beta}$ 
satisfies by construction the WI of Eq.(\ref{ward1}). Specifically, when  
the ``longitudinal'' part $L^{\mu\alpha\beta}$ is contracted with 
$q_{\mu}$ furnishes the r.h.s. of (\ref{ward1}), whereas
the ``transverse'' parts $T_1^{\mu\alpha\beta}$ and $T_2^{\mu\alpha\beta}$ are identically conserved.

({\bf iv})~ $\gv^{\mu\alpha\beta}$
contains  massless poles, i.e. terms going as $1/q^2$; as is well-known~\cite{Jackiw:1973tr}, 
the presence of such  poles 
allows for the possibility  $\Delta^{-1}(0) \neq 0$.
Note that not only the longitudinal but also one of the transverse parts 
contains such poles; this particular feature is motivated by the 
the one-loop analysis of the conventional three-gluon vertex 
presented in \cite{Ball:1980ax}.

({\bf   v})~   Clearly,    the   Lorentz   structures   appearing   in
(\ref{gtvertex}) do not  exhaust all tensorial possibilities.  Indeed,
the  most  general parametrization  of  a  vertex  with three  Lorentz
indices and  two independent momenta contains  14 linearly independent
tensors~\cite{Ball:1980ax,Binger:2006sj};  their  number may  be  reduced to  some
extent by  imposing Bose symmetry (only  partial in our  case) and the
constraint of the corresponding WI.  The simplified vertex proposed in
(\ref{gtvertex}) is only meant to capture some of the salient features
expected from  the full answer,  most importantly the correct  WI, the
Bose symmetry, and the presence of poles terms.

({\bf vi})~
The transverse  parts of $\gv^{\mu\alpha\beta}$ are  multiplied by the
constants $c_1$ and $c_2$, which, at this level of approximation, will
be treated  as arbitrary  adjustable parameters.  These  constants
affect not only the integral  equation but also  the  two  crucial
boundary conditions [(\ref{basreg1}) and (\ref{MC2})].  As we will see
in   detail  in   the  next   sections,  the   existence  or   not  of
self-consistent solutions for the  SDE (i.e.  solutions satisfying all
necessary  constraints) depends  crucially  on the  values chosen  for
these  constants. At first  sight this  seems to  run contrary  to the
standard  lore  of  the   gauge  technique,  according  to  which  the
transverse  parts in  the Ans\"atze  for  the vertices  do not  affect
appreciably the  solutions in the  IR.  Notice, however, that,  in the
case we consider  here, the distinction between what is  of UV and what
of IR  origin is not  so sharp.  Indeed,  as we will see  shortly, the
value  of  $\Delta^{-1}(0)$ -- clearly  an  IR quantity --  is  determined  by
integrals of the full propagator together with 
the running mass and coupling
over the  entire range of momenta; there  the UV behavior of
these quantities is essential.
For example, the running of the
masses  in the  UV is  controlled  by the  anomalous dimensions,  which
depend  themselves on  $c_1$  and/or $c_2$.   To  be sure,  in a  more
complete treatment, the  values of $c_1$ and $c_2$  should be uniquely
determined by the QCD dynamics, in particular by the SDE governing the
vertex $\gv^{\mu\alpha\beta}$;  unfortunately,  such an  analysis  is
beyond our powers at the moment.

We next write (\ref{polar2}) in the Euclidean space; 
to that end, we set $-q^2= q^2_{\da}$,
define $ \de(q^2_{\da})= - {\Delta}(-q^2_{\da})$,
and for the integration measure we have  $[dk] = i[dk]_{\chic E}= i d^4 k_{\chic E}/(2\pi)^4 $.
Then, using for ${\g}_{\nu\alpha\beta}$ the Ansatz of (\ref{gtvertex}), 
and suppressing the subscript ``E'',
(\ref{polar2}) becomes
\bea
\Delta^{-1}(q^2)  &=& q^2 - \frac{6 \tilde{b}g^2}{5\pi^2}\left[ \,\int d^4 k
\left(q^2 + \frac{2}{3} \bigg[k^2 - \frac{(k \cdot q)^2}{q^2}\bigg]\right)\Delta(k) \Delta(k+q)
- \int d^4 k \Delta(k)\right] \nonumber\\
&-& 
\frac{6 \tilde{b}g^2}{5\pi^2} c_1 \left[ \int d^4 k \, k^2 \, \Delta(k) \Delta(k+q)
-\int d^4 k  \Delta(k)\right]\nonumber\\
&-& \frac{6 \tilde{b}g^2}{5\pi^2} c_2 q^2 \int d^4 k \left[\Delta(k) - \Delta_0(k)\right]\Delta(k+q)\,,
\label{sde2}
\eea
where
$\tilde{b} \equiv 10 C_A/48\pi^2$ is the contribution of the gluons to the one-loop $\beta$ function 
in the PT-BFM scheme; 
the discrepancy from the correct value  $b= 11 C_A/48\pi^2$ is due to the 
omission of the ghosts~\cite{Abbott:1980hw,Aguilar:2006gr}.

The measure in  spherical coordinates is given by 
\be
\int d^{4} k = 
2 \pi\!\!\int_{0}^{\pi} \!\!\! d\chi\sin^2\chi\, 
\int_{0}^{\infty}\!\!\! dy y \, ;
\label{spher}
\ee 
introducing $q^2 \equiv x$, $k^2 \equiv y$ and $(k+q)^2 \equiv z$,
we have that $k \cdot q =\sqrt{xy}\cos\chi$, and so $(k \cdot q)^2/q^2 =y\cos^2\chi $, and 
$z = x + y +2 \sqrt{xy}\cos\chi$.
Due to the dependence of the unknown function on $\Delta(z)$, 
to solve (\ref{sde2}) one should carry out (numerically) the two integrals defined in (\ref{spher}).
Alternatively, one resorts to standard approximations (see (\ref{angle})) 
for the integral over $d\chi$, thus reducing the numerics to a single one-dimensional integral.
In our analysis we will use a modified version of the usual angular approximation
(see Appendix). 

The first important issue is the 
value that (\ref{sde2}) furnishes for $\Delta^{-1}(0)$;
in particular, in order to obtain IR-finite solutions,  $\Delta^{-1}(0)$ should be non-vanishing. 
It turns out that $\Delta^{-1}(0)$ may be 
determined from (\ref{sde2}) exactly, {\it before} doing any approximation.
Specifically, 
\bea
\int d^4 k \frac{(k \cdot q)^2}{q^2} \Delta(k) \Delta(k+q)\bigg|_{q^2\to 0}
&=& 2 \pi\! \!\int_{0}^{\pi} d\chi\sin^2\chi \cos^2\chi 
\int_{0}^{\infty}\!\!\! dy y^2 \Delta^2(y)
\nonumber\\
&=& \frac{1}{4} \int d^4 k \,k^2 \,\Delta^2(k)\,.
\eea
Thus, one obtains from (\ref{sde2})
\be
\Delta^{-1}(0) = \frac{3 \tilde{b}g^2}{5\pi^2} 
\Bigg[2(1+c_1) \int  d^4 k \,\Delta(k) - (1+2c_1)\int\,d^4 k \, k^2 \,\Delta^2(k)\Bigg]\,.
\label{Delta0}
\ee
Since the r.h.s. of (\ref{Delta0}) is divergent, in the next section 
this expression will be 
appropriately regularized \cite{Aguilar:2006gr}, 
following the rules of dimensional regularization;
there it will become clear 
why the UV running of the effective mass is of central importance.

Having determined the exact expression $\Delta^{-1}(0)$, one might be tempted 
to carry out the angular integration 
by employing the usual approximation of (\ref{angle}); however, particular
care is needed, since the straightforward application of (\ref{angle}) would 
misrepresent quantitatively some 
essential features of the original Eq.(\ref{sde2}),
and most importantly the terms determining the running of the mass.    
To remedy this, in the Appendix 
we implement judicious modifications to some of the relevant results of the 
angular approximation.

Let us denote the r.h.s. of (\ref{sde2}) by $I(q^2)$. 
Then we first write (\ref{sde2})  schematically in the form
\bea
\Delta^{-1}(x) &=& [I(x)-I(0)] + \Delta^{-1}(0)\nonumber\\
&\approx& [I_{{\chic M}{\chic A}}(x)-I_{{\chic M}{\chic A}}(0)] + \Delta^{-1}(0)\,,
\eea
where the exact $\Delta^{-1}(0)$ is given by (\ref{Delta0}), and the subscript ``MA''
denotes that the  (modified) angular approximation has been employed.
Then, using  for all terms appearing in (\ref{sde2})
the expressions given  in (\ref{I0}) and (\ref{mang}),
after some algebra we arrive at the SDE
\be
\Delta^{-1}(x) = Kx + \tilde{b}g^2 \sum_{i=1}^8 A_i(x)
+ \Delta^{-1}(0)\,, 
\label{sde}
\ee
with
\bea
A_1(x) &=& - \left(1+\frac{6 c_2}{5}\right) x \int_{x}^{\infty}\!\!\! dy\, y \Delta^{2}(y)\,,  
\nonumber\\ 
A_2(x) &=& \frac{6 c_2}{5} \,x \int_{x}^{\infty}\!\!\! dy  \Delta(y)\,,
\nonumber\\ 
A_3(x) &=& -\left(1+\frac{6 c_2}{5}  \right) x \Delta(x)\int_{0}^{x} \!\!\!dy \,y \Delta(y)\,,
\nonumber\\
A_4(x) &=& \left(-\frac{1}{10} -  \frac{3c_2}{5} + \frac{3c_1}{5}\right)
\int_{0}^{x}\!\!\! dy\, y^2 \Delta^2(y)\,,
\nonumber\\
A_5(x) &=& - \frac{6}{5}\bigg(1+ c_1\bigg) \Delta(x)\int_{0}^{x}\!\!\! dy\, y^2 \Delta(y)\,,
\nonumber\\
A_6(x) &=& \frac{6c_2 }{5} \int_{0}^{x}\!\!\! dy\, y \Delta(y)\,,
\nonumber\\
A_7(x) &=& \frac{2}{5}\, \frac{\Delta(x)}{x}\int_{0}^{x}\!\!\! dy\, y^3 \Delta(y)\,,
\nonumber\\
A_8(x) &=& \frac{1}{5x} \int_{0}^{x}\!\!\!dy\, y^3 \Delta^{2}(y)\,,
\label{Ai}
\eea
and $K$ is the wave-function renormalization constant, whose value is fixed by the  
renormalization condition 
${\Delta}^{-1}(\mu^2) = \mu^2$, with $\mu^2 \gg \Lambda^2$,  
where $\Lambda$ is the mass-scale of QCD. 

As has been discussed extensively in the literature,
due to the Abelian WIs satisfied by the PT effective Green's functions, 
$\Delta^{-1}(q^2)$ absorbs all  
the RG-logs~\cite{Cornwall:1981zr,Cornwall:1989gv,Binosi:2002ft,Binosi:2002vk},
exactly as happens in QED with the photon self-energy.
Equivalently, since $Z_{g}$ and $Z_{\widehat A}$, the renormalization constants 
of the gauge-coupling and the gluon self-energy, respectively, 
satisfy the QED relation ${Z}_{g} = Z^{-1/2}_{\widehat A}$~\cite{Abbott:1980hw}, 
the product 
$d(q^2) = g^2(\mu^2) \Delta(q^2,\mu^2)$ forms a RG-invariant 
($\mu$-independent) quantity;
for large momenta $q^2$,
\be
d(q^2) = \frac{g_{pert}^2(q^2)}{q^2}\,,
\label{ddef1}
\ee
where $g_{pert}^2(q^2)$ is the perturbative limit of the 
RG-invariant effective charge of QCD, i.e.
\be
g_{pert}^2(q^2) = \frac{g^2(\mu^2)}{1+  b g^2(\mu^2)\ln\left(q^2/\mu^2\right)}
= \frac{1}{b\ln\left(q^2/\Lambda^2\right)}\,.
\label{effch}
\ee
Notice however that Eq.(\ref{sde}) does not encode the correct RG behavior, in the sense that
it cannot be cast in a form containing 
only RG invariant quantities, such as  
$d(q^2)$ and $g_{pert}^2(q^2)$. This can be ultimately traced back to
the
fact that the various gauge technique inspired Ans\"atze for 
the full vertices (in our case the three-gluon vertex)
tend to 
mishandle their transverse (identically conserved parts); this, in turn, 
forces one to renormalize subtractively instead of multiplicatively.
We emphasize that there exist systematic methods for the 
construction of transverse parts with the correct UV properties~\cite{King:1982mk}, 
but have not been extended to the case of the three-gluon vertex.
Therefore, 
to restore the correct RGI properties, we follow  the heuristic procedure first proposed in 
\cite{Cornwall:1981zr,Cornwall:1985bg},
used recently also in ~\cite{Aguilar:2006gr}.
Specifically, 
every  $\Delta(t)$ ($t=y,x$) appearing inside  
the integrals on the r.h.s. of (\ref{sde}) is to be multiplied (by hand) by a factor 
$1+ {\tilde b} g^2\ln(t/\mu^2) = g^2 /g_{pert}^{2}(t)$, 
i.e. we carry out the replacement 
$\Delta(t) \rightarrow\left[g^2/ g_{pert}^{2}(t) \right]\Delta(t)$ (with $b\to{\tilde b}$). 
Then, one may rewrite (\ref{sde}) in terms of two  RG invariant quantities, 
$d(q^2)$, and
\be
{\cal L}(q^2) \equiv g_{pert}^{-2}(q^2) = \tilde{b}\ln\left(q^2/\Lambda^2\right)\,, 
\ee
as follows:
\be
d^{-1}(x) = K^{\prime}x + \tilde{b} \sum_{i=1}^8 {\widehat A}_i(x)\, + d^{-1}(0)\,,
\label{rgisde}
\ee
with
\bea
{\widehat A}_1(x) &=& - \left(1+\frac{6 c_2}{5}\right) x \int_{x}^{\infty}\!\!\! dy \,y\, {\cal L}^2(y) d^{\,2}(y)\,,   
\nonumber\\ 
{\widehat A}_2(x) &=& \frac{6 c_2}{5} x \int_{x}^{\infty}\!\!\! dy\, {\cal L}(y) d(y)\,, 
\nonumber\\ 
{\widehat A}_3(x) &=& -\left(1+\frac{6 c_2}{5}  \right) x\, {\cal L}(x) d(x)\int_{0}^{x}\!\!\! dy\, y\, {\cal L}(y) d(y)\,, 
\nonumber\\
{\widehat A}_4(x) &=& \left(-\frac{1}{10} -  \frac{3c_2}{5} + \frac{3c_1}{5}\right)
\int_{0}^{x}\!\!\! dy\, y^2\, {\cal L}^2(y) d^{\,2}(y)\,, 
\nonumber\\
{\widehat A}_5(x) &=& - \frac{6}{5}\bigg(1+ c_1\bigg)  {\cal L}(x) d(x)\int_{0}^{x}\!\!\! dy \,y^2\,{\cal L}(y) d(y)\,, 
\nonumber\\
{\widehat A}_6(x) &=& \frac{6c_2 }{5} \int_{0}^{x}\!\!\! dy\, y \,{\cal L}(y) d(y)\,, 
\nonumber\\
{\widehat A}_7(x) &=& \frac{2}{5}\, {\cal L}(x)\,\frac{d(x)}{x}\int_{0}^{x}\!\!\! dy\, y^3\,  {\cal L}(y) d(y)\,, 
\nonumber\\
{\widehat A}_8(x) &=& \frac{1}{5x} \int_{0}^{x} \!\!\! dy\, y^3\,  {\cal L}^2(y) d^{\,2}(y)\,, 
\label{hatAi}
\eea
where $ K^{\prime}= K/g^2$, given in closed form by
\be
K^{\prime} = 1 -  \frac{\tilde{b}}{\mu^2} \sum_{i=1}^8 {\widehat A}_i(\mu^2)\,,
\label{rconstant}
\ee
and
\be
d^{\,-1}(0) = \frac{3 \tilde{b}}{5\pi^2} 
\Bigg[2(1+c_1) \overline{{\cal T}}_{0} - (1+2c_1) \overline{{\cal T}}_{1}\Bigg]\,,
\label{D0}
\ee
with
\bea
\overline{{\cal T}}_{0} &=& \int  d^4 k \,{\cal L}(k^2)\, d(k^2)\,,
\nonumber\\
\overline{{\cal T}}_{1} &=& \int\,d^4 k \, k^2 \,{\cal L}^2(k^2)\,d^{\,2}(k^2)\,.
\eea

\setcounter{equation}{0}
\section{The UV behavior of the effective gluon mass}
\label{Sect:UVMASS}

In  this section  we will  discuss in  detail the  UV behavior  of the
effective gluon mass  obtained from the SDE of  (\ref{rgisde}).  As we
will see,  depending on the values  of the parameters  $c_1$ and $c_2$,
originally appearing  in the  vertex Ansatz of (\ref{LT1T2}),  one obtains
logarithmic or power-law running as two distinct dynamical possibilities.

We begin by briefly reviewing the importance of the running of the gluon mass
for obtaining a finite value for $d^{\,-1}(0)$; the finiteness of
$d^{\,-1}(0)$ is intimately linked to the renormalizability of the theory;
if $d^{\,-1}(0)$ turned out to be divergent, there would be 
no consistent way to eliminate this divergence by 
absorbing it into the renormalization of 
the parameters appearing in the fundamental QCD Lagrangian. 

As has been explained in \cite{Aguilar:2006gr}, 
under special assumptions on the form of $d(k^2)$, 
the r.h.s. of (\ref{D0})  
can be made finite by 
simply employing standard dimensional regularization results.
Specifically, let us
write $d(q^2)$ as 
\be
d(q^2) = g^2(q^2) \tilde\Delta(q^2)\,,
\label{ddef}
\ee
where 
\be
\tilde\Delta(q^2) =  \frac{1}{q^2 + m^2(q^2)}\,,
\label{tildeD}
\ee
and
\be
g^2(q^2) = \bigg[ \tilde{b}\ln\left(\frac{q^2 + f(q^2, m^2(q^2))}{\Lambda^2}\right)\bigg]^{-1}\,.
\label{GNP}
\ee
For large values of $q^2$, $\tilde\Delta(q^2)\to 1/q^2$ and $g^2(q^2)\to g^2_{pert}(q^2)$, 
and $d(q^2)$ reduces to its perturbative expression of (\ref{ddef1}). 
Then, it is relatively straightforward to show, 
using the elementary result
\be
\int  \frac{dz}{z\, (\ln z)^{1+\gamma}} = - \frac{1}{\gamma \, (\ln z)^{\gamma}}\,,
\label{elint}
\ee
that the difference 
$\overline{{\cal T}}_{0}-\overline{{\cal T}}_{1}$,   
given by 
\bea
\overline{{\cal T}}_{0}-\overline{{\cal T}}_{1} &=& 
\int\! d^4k\, {\cal L}^2(k^2)\, m^2(k^2) \,d^{\,2}(k^2) \nonumber \\
&&+ \, \tilde{b}\! \int\!  d^4k \, {\cal L}(k^2)\,g^2(k^2)\,
{d}(k^2)
\ln\left(1 + \frac{f(k^2,m^2(k^2))}{k^2}\right)\,,
\label{T0T1}
\eea
is finite, provided that  
$m^2(k^2)$ drops asymptotically at least as fast as 
$\ln^{-a} (k^2)$, with $a >1$, and 
$f(k^2, m^2(k^2))$ as  $\ln^{-c} (k^2)$, with $c>0$. 

Then, $\overline{{\cal T}}_{0}$ may be regularized simply by subtracting 
from it $\int\,d^4 k / k^2 = 0$, i.e.
\bea
\overline{{\cal T}}_{0}^{\,{{\rm reg}}} &=&  
\int\! d^4k \bigg(\,{\cal L}(k^2)\, d(k^2) -\frac{1}{k^2}\bigg)
\nonumber\\
&=& - \int\! d^4k \frac{m^2(k^2)\,\tilde\Delta(k^2)}{k^2}\,
- \tilde{b} \int\! d^4k\,\, d(k^2)\,
\ln\left(1 + \frac{f(k^2,m^2(k^2))}{k^2}\right)\,.
\label{basreg}
\eea

Thus, the regularized expression for $d^{\,-1}(0)$ is given by
\be
d^{\,-1}_{{{\rm reg}}}(0) = 
\frac{3 \tilde{b}}{5\pi^2}\Bigg[
\overline{{\cal T}}_{0}^{\,{{\rm reg}}} 
+ (1+2c_1) (\overline{{\cal T}}_{0}-\overline{{\cal T}}_{1})\Bigg]\,.
\label{basreg1}
\ee
When solving (\ref{rgisde}) one must impose (\ref{basreg1}) 
as an additional constraint; we will refer to (\ref{basreg1}) as the 
``seagull-condition''. The way this is done is by first choosing an 
arbitrary finite value for  the $d^{\,-1}(0)$ appearing in (\ref{rgisde})
and then solving it numerically. The solution obtained for $d(q^2)$ 
must be first decomposed following (\ref{ddef})--(\ref{GNP}), and then be substituted into 
the r.h.s. of (\ref{basreg1}); self-consistency requires that the resulting expression for 
$d^{\,-1}_{{{\rm reg}}}(0)$ must coincide with the initial value 
$d^{\,-1}(0)$. Evidently, the way that $m^2(q^2)$ runs is essential for 
this procedure, and can affect considerably the quantitative predictions.

In order to obtain from (\ref{rgisde}) the equation that determines the 
behavior of $m^2(x)$
at large $x$, first set in the r.h.s. of (\ref{hatAi}) 
$x{\cal L}(x)d(x)\to 1$,  ${\cal L}(x)d(x)\to 1/x$, and 
${\cal L}(y)d(y) = \tilde\Delta(y)$. Next, use the identity
$y\tilde\Delta(y) = 1 - m^2(y)\tilde\Delta(y)$ in all ${\widehat A}_i(x)$, 
keeping only terms linear in $m^2$ 
(terms quadratic in $m^2$ are subleading and may be safely neglected).
Then separate all contributions that go like $x$ from those that go like 
$m^2$ on both sides, and match them up. This gives rise to two independent 
equations, one for the ``kinetic'' term, which simply  
reproduces the asymptotic behavior $x\ln x$ on both sides, 
and one for the terms with $m^2(x)$, given by
\bea
m^2(x)\ln x &=& \tilde{b}^{-1} d^{-1}(0) +
a_1 \int_{0}^{x} dy \,m^2(y) \tilde\Delta(y) 
+ \frac{a_2}{x} \int_{0}^{x} dy \,y \,m^2(y) \tilde\Delta(y) 
\nonumber\\
&+&  \frac{a_3}{x^2}\int_{0}^{x} dy \,y^2 m^2(y) \tilde\Delta(y)
+ a_4 x \int_{x}^{\infty} dy\, m^2(y) \tilde\Delta^2(y)\,,
\label{meq}
\eea
with
\be
a_1 = \frac{6}{5} (1+c_2-c_1) \,,\,\,\,\,\,\,
a_2 = \frac{4}{5} + \frac{6c_1}{5} \,,\,\,\,\,\,\,
a_3 = - \frac{2}{5}\,, \,\,\,\,\,\,
a_4 = 1+ \frac{6c_2}{5} \,.
\label{ai}
\ee
Then, rewrite the first integral on the r.h.s of (\ref{meq2}) as
$\int_{0}^{x} = \int_{0}^{\infty} - \int_{x}^{\infty}$
to obtain 
\bea
m^2(x)\ln x  &=& {\cal C}
-a_1 \int_{x}^{\infty} dy \,m^2(y) \tilde\Delta(y) 
+ \frac{a_2}{x} \int_{0}^{x} dy \,y\, m^2(y) \tilde\Delta(y)\,\, \nonumber\\
&&+  \frac{a_3}{x^2}\int_{0}^{x} dy \,y^2 m^2(y) \tilde\Delta(y)
+ a_4 x \int_{x}^{\infty} dy\, m^2(y) \tilde\Delta^2(y)\,,
\label{meq2}
\eea
with
\be
{\cal C} \equiv  \tilde{b}^{-1} d^{-1}(0) + a_1 \int_{0}^{\infty} dy \,m^2(y) \tilde\Delta(y)\,.
\label{C0}
\ee

As we will see in a moment, the two 
possible asymptotic solutions of physical interest for $m^2(x)$  
are given by
\bea
m_1^2(x)  &=&  \lambda_1^2 (\ln x)^{-(1+\gamma_1)}\,,  
\label{m1}\\ 
m_2^2(x)  &=& \frac{\lambda_2^4}{x} (\ln x)^{\gamma_2-1}\,, 
\label{m2}
\eea
where $\lambda_1$ and $\lambda_2$ are two mass-scales,
and $\gamma_i >0$, $i=1,2$. 
The possibility of power-law running for the effective 
gluon mass, as expressed by (\ref{m2}), 
was first conjectured in \cite{Cornwall:1981zr,Cornwall:1985bg}, 
motivated by similar results in the study of chiral symmetry breaking.
Indeed, notice the similarity between the solutions given 
in (\ref{m1}) and (\ref{m2}) for the effective gluon mass 
and those appearing in the more familiar context of
the SDE (gap equation) for the quark self-energy; there,  
one finds the following two asymptotic solutions for 
the dynamically generated quark mass $M(x)$:
\bea
M_1 (x)  &=&  \mu_1 (\ln x)^{-\gamma_f}\,, \nonumber \\ 
M_2(x)  &=&  \frac{\mu_2^3}{x} (\ln x)^{\gamma_f-1}\,, 
\label{M12}
\eea
where 
$\gamma_f = 3C_f/16\pi^2 b$, with $C_f$ the Casimir eigenvalue 
of the fundamental representation [$C_f = (N^2-1)/2N$ for $SU(N)$]~\cite{com2}. 

Now, the important point to appreciate is that, in order for 
(\ref{meq2}) to have solutions vanishing in the UV, it is necessary that 
the constant term on the r.h.s. vanishes, i.e. ${\cal C}=0$.
If, for some reason, this condition cannot be implemented, the 
solution obtained will reach a constant value in the deep UV, thus invalidating
the basic characteristic of the dynamically generated mass.
Given that both $d^{-1}(0)$ and the integral appearing in (\ref{C0})
are manifestly positive quantities, one obvious necessary condition for 
obtaining ${\cal C}=0$ is that $a_1 < 0$. This requirement, in turn, 
restricts the possible values of the parameters $c_1$ and $c_2$, through the 
equation defining $a_1$ (first in (\ref{ai})). Assuming the correct sign for $a_1$, 
the way to actually enforce ${\cal C}=0$ will be completely dynamical: one must look
for masses with the appropriate momentum dependence such 
that the r.h.s. of (\ref{C0}) can be made equal to zero.
As we will see in the next section, 
the condition (\ref{C0}) [or, its improved version, (\ref{MC2})]
constrains the behavior of the dynamical mass in the IR and intermediate
momentum regimes.


Next, we set ${\cal C}=0$ in (\ref{meq2}) and verify that indeed $m_1(x)$ or $m_2(x)$ 
[Eqs.(\ref{m1}) and (\ref{m2})]
satisfy it. The upshot of this analysis will be that when $m_1(x)$ is substituted 
into the r.h.s of (\ref{meq2}) the first integral 
provides the solution, while all others are subleading,
 whereas for  $m_2(x)$ the leading contribution comes from the 
second integral, and the other three are subleading.
(The third and fourth integrals are thus 
subleading for both types of solutions).

In our demonstration we will employ the asymptotic property of the 
incomplete $\Gamma$ function. The latter is defined as \cite{GR}
\be
\Gamma(\alpha,u) = \int_{u}^{\infty} dt\, e^{-t}\, t^{\alpha -1}\,, 
\label{Gdef}
\ee
(with no restriction on the sign of $\alpha$),
and its asymptotic representation for large values of $|u|$ is given by
\be
\Gamma(\alpha,u) = u^{\alpha -1} e^{-u} + {\cal O}(|u|^{-1})\,.
\label{Gas}
\ee

To see in detail what happens in the case of the logarithmic running, 
substitute (\ref{m1}) into both sides of (\ref{meq2})
and use that asymptotically $\tilde\Delta(y)\to y^{-1}$. Then 
\bea
\int_{x}^{\infty} dy \,\frac{m_1^2(y)}{y} &=&  \gamma_1^{-1} \, m_1^2(x) \,\ln x \,,
\label{intm11}
\\ 
\frac{1}{x}\int_{0}^{x} dy \,m_1^2(y) &=& m_1^2(x) + {\cal O}\left( 1/\ln x\right) \,,
\label{intm12}
\\ 
\frac{1}{x^2} \int_{0}^{x} dy \,y m_1^2(y) &=& \frac{m_1^2(x)}{2}  +  {\cal O}\left( 1/\ln x\right) \,,
\label{intm13}
\\ 
x \int_{x}^{\infty} dy\, \frac{m_1^2(y)}{y^2} &=& m_1^2(x) + {\cal O}\left( 1/\ln x\right)\,. 
\label{intm14}
\eea
In evaluating the integral of (\ref{intm11}) 
we have used (\ref{elint}); for the remaining three we  
have set $y = e^{-t}$ in (\ref{intm12}), $y = e^{-t/2}$ in (\ref{intm13}),
and $y = e^{t}$ in (\ref{intm14}), to cast them
into the form of the incomplete $\Gamma$ function given in (\ref{Gdef}), 
and have subsequently used the
asymptotic expression of (\ref{Gas}).  

Thus, for asymptotic values of $x$ the dominant contribution comes from (\ref{intm11}) 
Substituting into (\ref{meq2}), we see that both sides 
can be made equal provided that 
\be
\gamma_1=-a_1 \,.
\label{g1a1}
\ee
Since $\gamma_1$ must be positive, it follows that $a_1 < 0$, consistent with the 
requirement imposed by Eq.~(\ref{C0}), as discussed after  Eq.~(\ref{M12}).

Let us next turn to the case of power running, substituting (\ref{m2}) into both sides of (\ref{meq2}).
Now the leading contribution comes from the integral proportional to $a_2$ in (\ref{meq2}).
Specifically,
\bea
\int_{x}^{\infty} dy \,\frac{m_2^2(y)}{y} &=&   m_2^2(x) + {\cal O}\left( 1/\ln x\right)\,,
\label{intm21}
\\ 
\frac{1}{x}\int_{0}^{x} dy \,m_2^2(y) &=& \gamma_2^{-1}   
m_2^2(x) \ln x + \frac{c^{\prime}}{x} \,,
\label{intm22}
\\
\frac{1}{x^2} \int_{0}^{x} dy \,y m_2^2(y) &=&  m_2^2(x)  + {\cal O}\left( 1/\ln x\right) \,,
\label{intm23}
\\ 
x \int_{x}^{\infty} dy\, \frac{m_2^2(y)}{y^2} &=& \frac{m_2^2(x)}{2} + {\cal O}\left( 1/\ln x\right)\,.
\label{intm24}
\eea
In evaluating (\ref{intm22}) we have used (\ref{elint}).
The constant $c^{\prime}$  comes from the lower limit of the integral; it
is finite, because in that limit one must use inside the integral 
the full $y \tilde\Delta (y)$, which is infrared safe due to the presence of the mass.
The term proportional to  $c^{\prime}$ is suppressed by a factor $\ln^{\gamma_2} x$
(assuming $\gamma_2 >0$) compared to the first term,
and can therefore be neglected. As in the case of the logarithmic running,
for the other three integrals we have used the appropriate change of variables to
cast them into the incomplete $\Gamma$ function, resorting again  
to its asymptotic expression.
Thus, we conclude that $m_2^2(x)$ satisfies  (\ref{meq2}) provided that 
\be
\gamma_2 = a_2 \,. 
\label{g2a2}
\ee
This condition, 
in turn, constrains the possible values of $c_1$ appearing in the definition of $a_2$ 
(second of (\ref{ai})). 

We emphasize again that, for either of the two 
physically relevant possibilities given by (\ref{m1}) and (\ref{m2})
to be realized,  the constant term on the l.h.s of (\ref{C0}) must be forced to
vanish, by imposing the mass-condition
\be
d^{-1}(0) = \gamma_1 \tilde{b} \int_{0}^{\infty} dy \,m^2(y) \tilde\Delta(y)\,.
\label{MC1}
\ee

In addition, it is important to mention 
that whereas the individual (\ref{m1}) and (\ref{m2})
are {\it separately} solutions of (\ref{meq2}), a linear combination of the form
$m^2(x) = C_1 m^2_1(x) + C_2 m^2_2(x)$
cannot be regarded as a solution. The reason   
is that for $m_1^2(x)$ to be a solution one neglects  
all terms in (\ref{intm12})-(\ref{intm14}), which are, however, 
clearly larger than $m_2^2(x)$. This does not necessarily mean that the two runnings cannot
coexist, it simply says that the possible coexistence cannot be self-consistently inferred
from (\ref{meq2}). For this reason, in the analysis of the 
next section the two possibilities will be treated separately.  
It should also be clear that 
the terms quadratic in $m_2^2(x)$ that have been dropped when 
deriving (\ref{meq}) are indeed subleading  
for both types of asymptotic behavior, (\ref{m1}) and (\ref{m2}).

Since in this section we have been mainly interested in the 
UV running of $m^2(x)$, in the analysis presented 
above we have used for the $g^2(q^2)$ of (\ref{GNP})
its perturbative expression, given in (\ref{effch}), 
i.e. we have replaced $g^2(q^2)\to g_{pert}^2(q^2)$.
A complete treatment, where the full expression for $g^2(q^2)$ is kept,
does not affect in the least the conclusions regarding the 
UV behavior of  $m^2(x)$, but modifies slightly the   
condition (\ref{MC1}) in the IR. Specifically, 
\be
d^{-1}(0) = (1+\gamma_1) \,\tilde{b}
\int_{0}^{\infty}\!\!\!\! dy \,y\,m^2(y)\,{\cal L}^{\,2}(y)\,d^{\,2}(y)
-\tilde{b}\int_{0}^{\infty}\!\!\!\!dy \,m^2(y)\,{\cal L}(y)\,d(y)\,.
\label{MC2}
\ee
Evidently, in the limit ${\cal L}(y)\,d(y)\to \tilde\Delta(y)$, (\ref{MC2}) reduces to (\ref{MC1}), 
as it should. In the numerical analysis that follows 
we will always use (\ref{MC2}), referring to it as the ``mass-condition''.


\setcounter{equation}{0}
\section{Numerical analysis} 
\label{Sect:NUM}

In this section we will solve numerically
the integral equation given in Eq.(\ref{rgisde})
supplemented by the renormalization condition (\ref{rconstant}),
and 
subject to the 
two constrains imposed by the mass- and seagull-conditions, 
Eqs.(\ref{MC2}) and (\ref{basreg1}), respectively. 
As mentioned already in the previous sections, and as we will see 
in detail in what follows,
the first condition restricts the momentum-dependence of the 
mass  in the  intermediate and deep infrared regimes, 
while the latter furnishes essentially
the  value of $d^{-1}(0)$.

When dealing with this problem we have at our disposal
three undetermined parameters: $c_1$ and $c_2$
appearing in the Ansatz for the three-gluon vertex 
[Eqs.(\ref{gtvertex}) and (\ref{LT1T2})], and the value of $d^{-1}(0)$.
It turns out that the 
simultaneous solution of the
integral equation and its constrains 
restricts considerably the acceptable combinations of these parameters.

The strategy we will employ in our numerical analysis consists of the following main steps:

({\bf i})~ We choose an arbitrary initial value for $d^{-1}(0)$, 
to be denoted by $d^{-1}_{in}(0)$, together with a set of values for 
$c_1$, $c_2$, and we substitute them into the 
integral  equation  (\ref{rgisde}), generating a solution  for  $d(q^2)$.

({\bf ii})~ 
The solution for $d(q^2)$ obtained in ({\bf i}) must be then  
decomposed 
as the product of $\tilde\Delta(q^2)$ and $\gnp(q^2)$, 
according to Eqs.(\ref{ddef}), (\ref{tildeD}), and (\ref{GNP}).
To do this, first a simple Ansatz for $m^2(q^2)$ 
is written down, which in the UV 
displays one of the two 
physically relevant asymptotic behaviors (logarithmic or power-law),
while in the IR reaches a finite value.
The generic form of these two types of Ans\"atze is given in 
(\ref{dmass_log}) and (\ref{dmass_power}), for logarithmic and power-law running, respectively.
The anomalous dimensions $\gamma_1$ and $\gamma_2$ 
appearing there are linear combinations of  $c_1$ and $c_2$  
[given by (\ref{ai}),(\ref{g1a1}), and (\ref{g2a2})], and 
control the behavior of $m^2(q^2)$ in the deep UV, whereas
the parameters $\rho$ and $m_0$ are free for the moment, and affect 
the intermediate and IR regions.  

({\bf iii})~
The integrals on the r.h.s. of the mass-condition (\ref{MC2}) are evaluated numerically, 
using as input 
the Ansatz for $m^2(q^2)$ chosen in ({\bf ii}), together with
the numerical solution for  $d(q^2)$; this furnishes a value for 
the $d^{-1}(0)$ on the l.h.s.
By varying  
$\rho$ and $m_0$ we try to make that $d^{-1}(0)$  
match  $d^{-1}_{in}(0)$ of step ({\bf i}). 
We may or may not be able to do this, depending on the values of  $c_1$ and $c_2$,
and the type of Ansatz (logarithmic or power-law) chosen for the mass. 
In general, in the cases where (\ref{MC2}) can be satisfied, one finds that  
this may be accomplished not just for one but 
for various sets of values for $\rho$ and $m_0$. This, in turn, gives rise 
to a family of possible masses, which have a common behavior in the deep UV, 
but differ in the intermediate and IR regions.

({\bf iv})~
Once the family of allowed masses has been determined from the mass-condition 
in (iii), one extracts from (\ref{ddef}) 
the corresponding families of non-perturbative 
effective charges $\gnp(q^2)$; one simply multiplies the numerical points of 
$d(q^2)$ by $[q^2+ m^2(q^2)]$.
On physical grounds we require that
the resulting effective charges should be {\it monotonically decreasing} 
functions of $q^2$; this requirement eliminates any  
member of the mass family that gives rise to effective charges with ``bumps''.
All $\gnp(q^2)$ so obtained display in the UV 
the logarithmic running expected from the one-loop RG (i.e. asymptotic freedom 
corresponding to \mbox{$\beta=-\tilde{b}g^3$}), and reach a finite value in the deep IR. 
For the implementation of the seagull-condition, (\ref{basreg1}), 
one must supply the functions $f(q^2,m^2(q^2))$, appearing in (\ref{GNP}). 
Therefore, we must extract for each $\gnp(q^2)$ the  corresponding $f(q^2,m^2(q^2))$; 
the way this is done is by fitting the numerical points
determining $\gnp(q^2)$ by Eq.~(\ref{GNP}), 
assuming for $f(q^2,m^2(q^2))$ the expression of (\ref{func_fit}).

({\bf v})~
We next substitute $\gnp(q^2)$, $m^2(q^2)$ and $d(q^2)$ into the
integrals on the r.h.s. of the seagull-condition (\ref{basreg1}),
whose value is computed numerically; this furnishes the value for  
$d^{-1}(0)$ appearing on the r.h.s.
If this value for $d^{-1}(0)$ does not coincide with  $d^{-1}_{in}(0)$
[we require an accuracy of about 1 part in $10^3$]
 a new set of values for 
$c_1$ and $c_2$ is chosen (keeping $d^{-1}_{in}(0)$ fixed), and the  
procedure is repeated from the beginning, until coincidence has been reached.
At that point we consider to have found a solution, namely 
the $d(q^2)$ obtained for $d^{-1}_{in}(0)$
and the values for $c_1$ and $c_2$ used the last (and only ``successful'') iteration.

({\bf vi})~
A different value for $d^{-1}_{in}(0)$ is chosen, and the procedure is repeated 
starting from step ({\bf i}).

To  solve  the  integral  equation  we  employ  a  simple  iterative
procedure, where an initial guess  is made for the solution $d(q^2)$ on
a     discretized     momentum    grid     in     the    domain     of
$[0,\Lambda_{\rm{UV}}]$.
More  specifically, the grid  is split in
two regions $[0,\mu^2]$  and $(\mu^2,\Lambda_{\rm{UV}}]$ whose purpose
is  allow  for the  implementation  of  the renormalization  condition,
given     by     Eq.~(\ref{rconstant}).     
Typically,     we     choose     $
\Lambda_{\rm{UV}}=10^{\,6}     \;\mbox{GeV}^{\,2}$,    $\mu^2=M^2_{\rm
Z}=(91.18)^2  \; \mbox{GeV}^{\,2}$  and we  used as  input a  value of
$\Lambda = 300 \;\mbox{MeV}$ for the QCD mass-scale.

Our numerical analysis reveals a clear separation between
the two types of 
asymptotic behavior for $m^2(q^2)$  depending on  the values chosen 
for $c_1$ and $c_2$.
Specifically, for \mbox{$c_1 \in [0.15,0.4]$} and $c_2 \in [-1.07, -0.92]$
the asymptotic  behavior of $m^2(q^2)$ is given  by Eq.(\ref{m1}),
whereas for $c_1 \in [0.7,1.3]$  and $c_2 \in [-1.35, -0.68]$
the $m^2(q^2)$  displays the 
power-law running  of Eq.(\ref{m2}).

\vspace{-0.7cm}

\subsection{$m^2(q^2)$ with logarithmic running}

\vspace{-0.5cm}

When solving Eq.(\ref{rgisde})  
choosing values for $c_1$ and $c_2$ from the intervals 
$c_1 \in [0.15,0.4]$ and $c_2 \in [-1.07, -0.92]$,
the constraints (\ref{MC2}) and (\ref{basreg1}) 
 can be simultaneously  satisfied, and the 
$d(q^2)$ obtained may indeed be decomposed as in Eq.(\ref{ddef}),
with a functional Ansatz 
for the running mass of the form
\be
m^2(q^2)=m^2_0\Bigg[\ln
\left(\frac{q^2+\rho\,m^2_0}{\Lambda^2}\right)\Big/\ln\left(\frac{\rho\,m^2_0}{\Lambda^2}\right) \Bigg]^{-1-\gamma_1} \,,
\label{dmass_log}
\ee 
where $\gamma_1=-a_1$  [see (\ref{g1a1}) and (\ref{ai})]. Evidently, for large $q^2$ the above expression 
goes over to the logarithmic  behavior described by Eq.(\ref{m1}),  
with $\lambda_1^2 = m^2_0 \left[\ln\left(\frac{\rho\,m^2_0}{\Lambda^2}\right)\right]^{1+\gamma_1}$.
This simple Ansatz connects continuously the UV and IR regions; 
at $q^2=0$ reaches the finite value $m^2(0)=m^2_0$.  

The  parameters $m_0$ and  $\rho$ appearing  in Eq.(\ref{dmass_log})
control the way the mass runs in the intermediate and IR regions; their
values are restricted by the mass-condition, Eq.(\ref{MC2}). To impose
the mass-condition,  we first choose  a random value for  $m_0$, and
then we search for values of $\rho$ that satisfy Eq.(\ref{MC2}).  Even
though this procedure does not single  out a unique pair of values for
$m_0$ and $\rho$, it restricts considerably their allowed range.  In
fact, the acceptable range for $(m_0,\rho)$ gets further restricted
by  imposing   the  additional  requirements   that  
all the $m^2(q^2)$ and  
the effective  charges 
generated subsequently from them 
(by multiplying  $d(q^2)$  by $[q^2+  m^2(q^2)]$) should  be
monotonically decreasing functions of $q^2$.
The combination of all these constraints leads eventually to rather stable results: 
if a pair $(m_0,\rho)$ furnishes a consistent solution  
for a given $(c_1,c_2)$ , then any  other pair 
$({m_0^{\prime}},\rho^{\prime})$ is also a solution, provided that 
$c_2$ is only slightly adjusted (less than $5\%$).

The running couplings $\gnp(q^2)$ obtained using Eq.(\ref{dmass_log}) in Eq.(\ref{ddef}) 
can be fitted
very accurately by means  of Eq.(\ref{GNP}), with the function $f(q^2,m^2(q^2))$ given by
\be
f(q^2, m^2(q^2)) = \rho_{\,1} m^2(q^2)+ \rho_{\,2} \frac{m^4(q^2)}{q^2+m^2(q^2)} +\rho_{\,3} \frac{m^6(q^2)}{[q^2+m^2(q^2)]^{\,2}} \,,
\label{func_fit}
\ee

\begin{figure}[ht]
\vspace{-0.5cm}
\hspace{-1.5cm}
\includegraphics[scale=2.0]{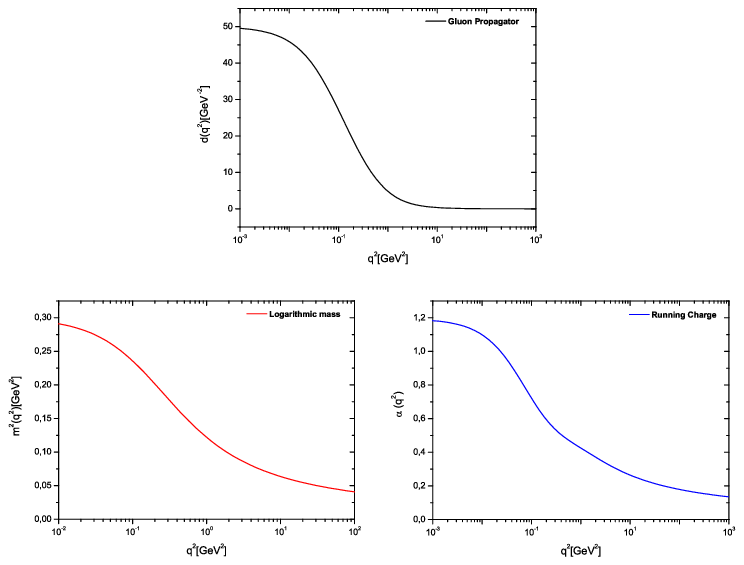}
\vspace{-0.5cm}
\caption{\small{Solutions obtained for the choice  $d^{\,-1}(0)=0.02  \;\mbox{GeV}^{\,2}$, $c_1=0.15$ and $c_2=-0.9635$.
 Upper panel: the numerical solution for $d(q^2)$. Lower panels: the logarithmic dynamical mass, $m^2(q^2)$, for $m_0^2=0.3 \;\mbox{GeV}^{\,2}$ and $\rho=1.007$ in the Eq.(\ref{dmass_log}). On the right panel we show  the running charge, $\alpha(q^2)=\gnp(q^2)/4\pi$, which can be fitted using Eqs.(\ref{GNP}) and (\ref{func_fit}) with $\rho_1=6.378$, $\rho_2=-8.984$ and $\rho_3=3.466$.}}
\label{fb}
\end{figure}

In  Fig.(\ref{fb}) we present a 
typical solution for $d(q^2)$, $m^2(q^2)$, and  the effective charge
\mbox{$\alpha(q^2)=\gnp(q^2)/4\pi$}, respectively.
%

\vspace{-0.5cm}
\subsection{$m^2(q^2)$ with power-law running}
\vspace{-0.5cm}

As  we increase  $c_1$,  Eq.(\ref{MC2})  can not  be  satisfied if  we
insist  on   imposing  the  logarithmic  running   for  $m^2(q^2)$.  In
particular, for $c_1 \in [0.7,1.3]$, $c_2 \in [-1.35, -0.68]$
and  \mbox{$d^{\,-1}(0) \in[0.01\;\mbox{GeV}^2,0.04  \;\mbox{GeV}^2]$} 
we have verified
that the mass-condition can be satisfied   only if   instead  of  
Eq.(\ref{dmass_log}) we use  the following  functional form  for the
running mass
\be
m^2(q^2)=\frac{m^4_0}{q^2+m^2_0}\Bigg[\ln
\left(\frac{q^2+\rho\,m^2_0}{\Lambda^2}\right)\Big/\ln\left(\frac{\rho\,m^2_0}{\Lambda^2}\right) \Bigg]^{\gamma_2-1} \,,
\label{dmass_power}
\ee 
with $\gamma_2= a_2$ [see (\ref{g2a2}) and (\ref{ai})].
Evidently, this Ansatz corresponds to power-law running for the mass;
for large $q^2$, the $m^2(q^2)$ goes over to the solution 
denoted by $m^2_2(q^2)$ in (\ref{m2}), with 
$\lambda_2^4 = m^4_0 \left[\ln\left(\frac{\rho\,m^2_0}{\Lambda^2}\right)\right]^{1-\gamma_2}$.
Clearly, (\ref{dmass_power}) is the simplest extension of 
the asymptotic expression (\ref{m2}) to the entire range of momenta,
from the UV all the way down to $q^2=0$, where it assumes the 
finite value $m^2(0)= m^2_0$.

As  in  the  previous  case, only those sets of $m_0$  and  $\rho$
that satisfy the mass-condition (\ref{MC2}) are allowed.
In fact, in the  case of the power-law running
this condition turns out to be significantly more restrictive 
than in the logarithmic case. 
The reason is that, due to  the faster 
decrease of the mass in the UV, the leading  contribution to the 
mass-condition comes now from the 
intermediate and IR regions; therefore, the result  
is much more sensitivity  to small variations of $m_0$  and  $\rho$. 

\begin{figure}[hb]
\hspace{-1.5cm}
\includegraphics[scale=2.0]{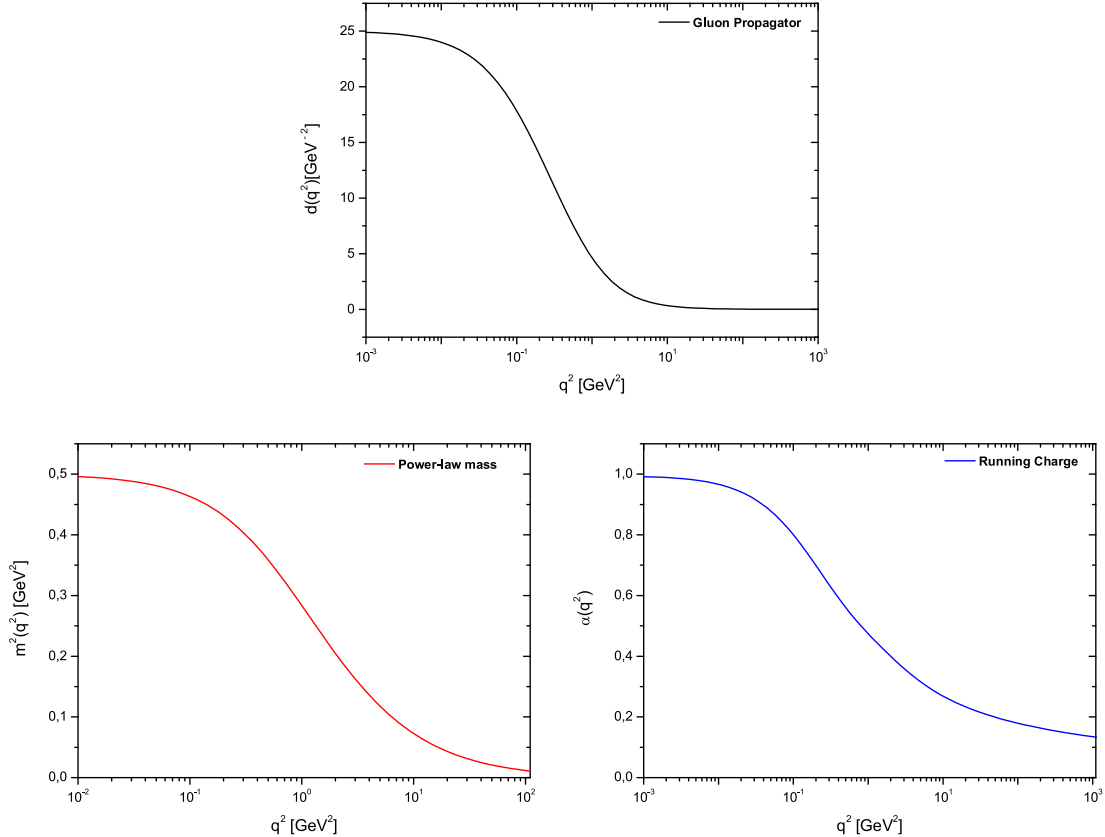}
\vspace{-0.5cm}
\caption{\small{Solutions obtained for the choice  $d^{\,-1}(0)=0.04  \;\mbox{GeV}^{\,2}$, $c_1=1.1$ and $c_2=-1.121$.
 Upper panel: the numerical solution for $d(q^2)$. Lower panels: the power-law dynamical mass, $m^2(q^2)$ for
$m_0^2=0.5 \;\mbox{GeV}^{\,2}$ and $\rho=1.046$ in the Eq.(\ref{dmass_power}). On the right panel we show the running charge, $\alpha(q^2)=\gnp(q^2)/4\pi$,  which can be fitted by Eqs.(\ref{GNP}) and (\ref{func_fit}) with
$\rho_1=1.205$, $\rho_2=-0.690$ and $\rho_3=0.121$.}}
\label{fc}
\end{figure}

A typical solution for a choice of  $c_1$, $c_2$, and  $d^{\,-1}(0)$
within the aforementioned ranges is shown in Fig.(\ref{fc}).
The $d(q^2)$ is decomposed according to (\ref{ddef})
into an $m^2(q^2)$ of the general form given  in
Eq.(\ref{dmass_power}) and an effective  charge  $\alpha(q^2)$; 
the latter is fitted using again (\ref{GNP}) and (\ref{func_fit}).
%

\vspace{5cm}

\begin{center}
\begin{figure}[ht]
\hspace{-1.2cm}
\includegraphics[scale=2.0]{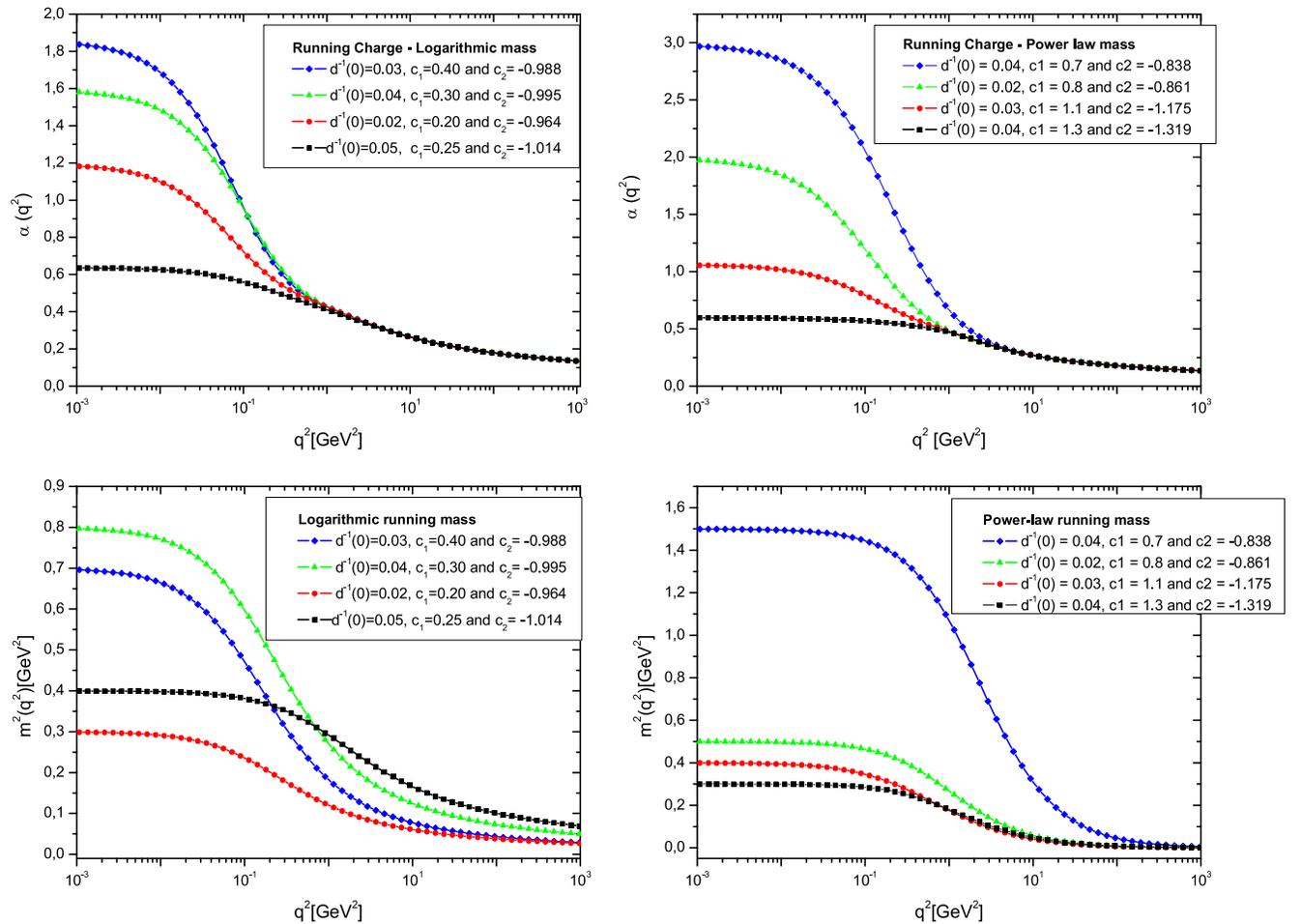}
\caption{Upper panels: The running charges, $\alpha(q^2)=\gnp(q^2)/4\pi$, corresponding to the different choices of 
$d^{-1}(0)$, $c_1$ and $c_2$ for the logarithmic running mass (left panel)
and for the power-law running mass (right panel). The corresponding masses are plotted in the lower panels.}
\label{fd}
\end{figure}
\end{center}

In the upper panels of Fig.(\ref{fd}) we  plot a  series of effective charges obtained by
fixing different set of values for $d^{\,-1}(0)$, $c_1$, and $c_2$. All
these  coupling   were  subjected   to  the  constraints   imposed  by
Eqs.(\ref{MC2})  and (\ref{basreg1}); the  corresponding values
for  $c_1$  and  $c_2$  are  given in  the  legend. 
The respective logarithmic and power-law masses are shown in the lower
panels. Observe  that
$\alpha(0)$ shows a  strong dependence on the values  chosen for 
$c_1$ and  $c_2$; this is so, even if the same  value for $d^{\,-1}(0)$ 
is chosen.
The difference between effective charges obtained with the
same  $d^{\,-1}(0)$ is due to the fact that one is forced  to  use different
values of  $m_0^2$ in order to  satisfy the mass-condition.
Thus, by changing the value of $m_0^2$, one obtains different values of $\alpha(0)$
for the same $d^{-1}(0)$.

\begin{figure}[hb]
\hspace{-1.5cm}
\includegraphics[scale=2.25]{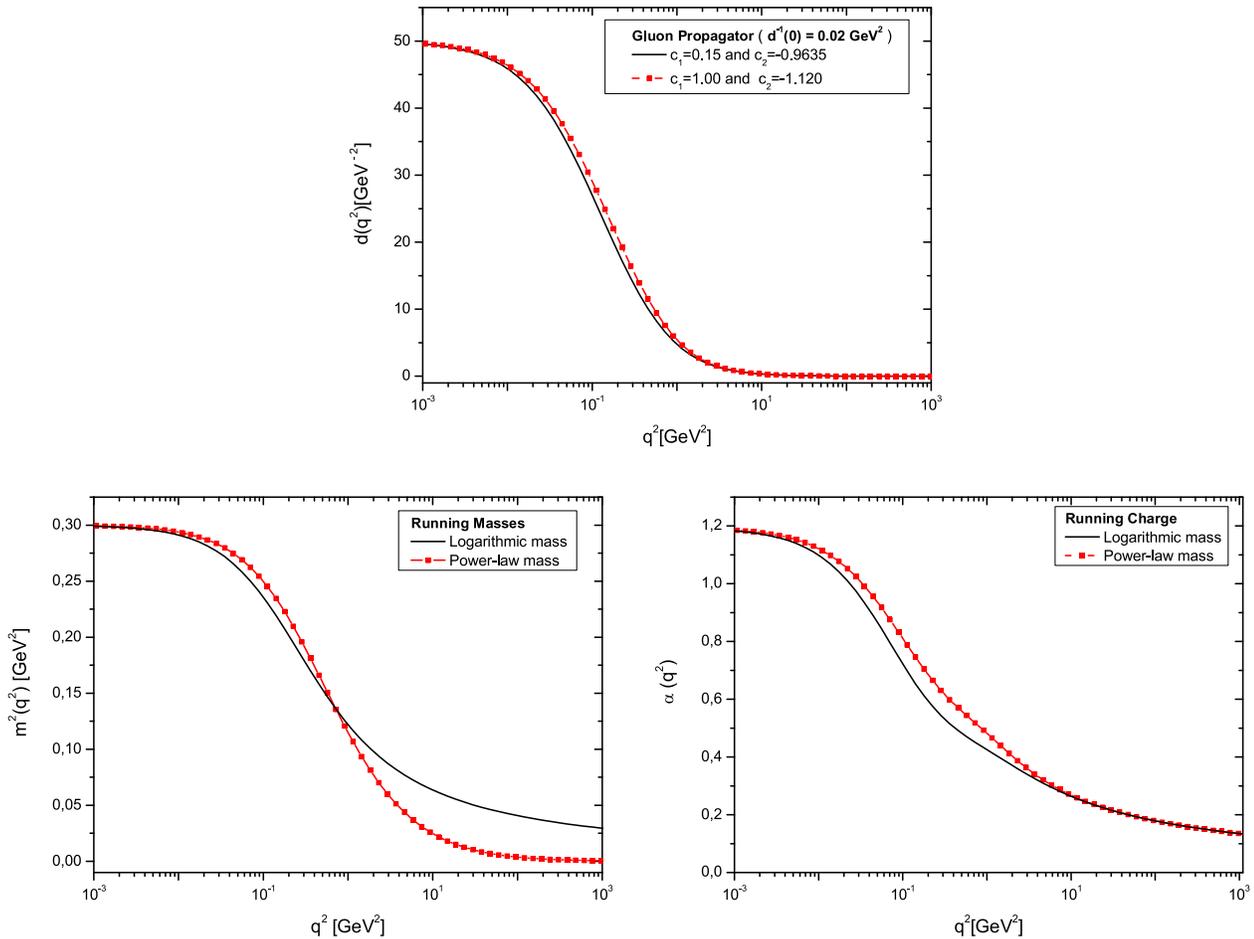}
\vspace{-0.5cm}
\caption{Comparison between logarithmic and a power-law running cases. 
Upper Panel: The numerical solutions for
$d(q^2)$, when  $d^{\,-1}(0)=0.02, \;\mbox{GeV}^{\,2}$, $c_1=0.15$ and $c_2=-0.9635$ (Logarithmic case) and for
$c_1=1.00$ and $c_2=-1.12$ (Power-law). Lower Panels: The corresponding logarithmic and a power-law running masses and the running charges.}
\label{fe}
\end{figure}

In Fig.(\ref{fe}) we show a comparison
between logarithmic and a power-law running, by choosing $c_1$ and $c_2$ from the
corresponding intervals, for fixed values of $d^{\,-1}(0)$ and $m_0^2$.
The corresponding $d(q^2)$ are shown in the upper panel, whereas the running 
masses and couplings are plotted in the lower left and right panels, respectively.
The faster decrease of the power-law running mass in the UV is clearly visible.
It is evident that, as already mentioned, 
in the power-law case the leading contribution to the mass-condition  comes
mainly from the IR and intermediate regions, while in the logarithmic case the UV region provides
a considerable support.
Since the UV behavior of the two $d(q^2)$ and $\alpha(q^2)$ is essentially 
fixed by asymptotic freedom, whereas their IR regimes are determined, to a large extent, by the value of $m_0^2$,
the difference between the two cases
is perceptible only in the intermediate momentum region.


\setcounter{equation}{0}
\section{Discussion and Conclusions}
\label{Sect:Concl}

In this article we have  studied both analytically and numerically the
non-linear  SDE for  the gluon  self-energy in  the  PT-BFM formalism,
focusing  on the  dynamical generation  of a  gauge-invariant infrared
cutoff.  In particular, we have established the existence of IR-finite
solutions,  i.e. solutions  that are  finite  in the  entire range  of
momenta, displaying asymptotic freedom in the UV and reaching a finite
positive  value at  $q^2=0$.   This non-perturbative  behavior may  be
described in  terms of an  effective gluon mass, whose  presence tames
the  perturbative Landau singularity  and prevents  $\Delta(q^2)$ from
diverging  in the  IR.  Just  as  happens with  the constituent  quark
masses,  this  dynamical  gluon  mass  depends  non-trivially  on  the
momentum.   Our study  of the  non-linear  SDE has  revealed that  the
dynamical  mass $m^2(q^2)$ may  have two  different types  of functional
dependence on $q^2$  in the deep UV: (i) $m^2(q^2)$  drops as an inverse
power  of a  logarithmic; this  behavior has  also been  found  in the
studies of  linearized SDE, and (ii) $m^2(q^2)$  with power-law running,
i.e. the  mass drops off as  $1/q^2$.  This type of  solution is found
for the first time  in the context of a SDE.  At  the level of the SDE
we  study  either  type   of  asymptotic  behavior  (logarithmic  and
power-law)  may   be  obtained,  depending  on  the   details  of  the
three-gluon vertex.   The latter depends on two  parameters, $c_1$ and
$c_2$,  which  control  the   relative  contribution  of  its  various
tensorial structures.  Our numerical analysis reveals that the sets of
values  for $c_1$  and $c_2$  that  give rise  to logarithmic  running
belong to  an interval that  is disjoint and well-separated  from that
producing power-law running.

The  possibility of gluon masses falling like the inverse square of the momentum    
has been anticipated~\cite{Cornwall:1981zr}
by analogy with  
the constituent quark masses generated from the standard 
gap equation for the quark self-energy
[second equation in (\ref{M12})].
In addition,  general OPE considerations support the  existence of a 
$m^2(q^2)$ displaying power-law running.
Assuming that  the  OPE  holds for  a  quantity like  
$m^2(q^2)$, and given  that, in the absence of quarks, 
 $\langle  G^2 \rangle$  is the
lowest order (dimension four) local gauge-invariant condensate,  
then one would  expect that asymptotically, and up to logarithms, 
$m^2(q^2)\sim \langle G^2  \rangle/q^2$, exactly as was found in~\cite{Lavelle:1991ve}.

To be sure, the various connections between the effective gluon mass and 
the OPE, the gluon condensates, and the QCD sum rules, 
deserve a detailed, in-depth study.
One issue is the type of modifications induced 
to the OPE predictions 
for observables (i.e. correlators of gauge-invariant currents) if one were to 
use in their calculation gluon propagators with 
a dynamical (or even hard) mass, as was first done in~\cite{Graziani:1984cs}.
In addition,   
despite important contributions   in   this
direction~\cite{Lavelle:1991ve,Kogan:1994wf,Gorbar:1999xi},           
a definite,  first-principle  relation between  $m^2$  and 
$\langle  G^2 \rangle$ (or  other condensates~\cite{amin}) still eludes us.
In the context of the SDE this is   
mainly because the CJT formalism~\cite{Cornwall:1974vz} 
has not been yet  fully adapted to  treat gluon mass
generation in a consistent way.  Qualitatively speaking, the CJT effective
potential $V$ is given by
\be V = -  \frac{1}{2} {\rm Tr} \ln \left(\Delta\Delta_0^{-1}\right) +
\frac{1}{2} \left({\rm  Tr}\Delta\Delta_0^{-1}\, - 1\right)  + V_{{\rm 2PI}}
\label{CJT}
\ee 
where  the trace  is taken in  the functional sense,  and 
$V_{{\rm 2PI}}$  denotes  the   contributions  from  the  (appropriately
dressed) two-particle irreducible graphs.  In the original formulation
$V$ is a functional of  the conventional gluon propagators, and higher
point Green functions; its extremization with respect to any of yields
the corresponding SDE's.  To make reliable contact with the results of
the BFM-PT, one should modify Eq.(\ref{CJT}) appropriately, expressing
it  in terms  of the  gauge-invariant PT  gluon propagator.  Thus, the
gluon mass will enter into  $V$ through the massive gluon propagators;
then, the minimization  of $V$ will yield  a theoretical expression
for the energy  density of the QCD vacuum, which must  be set equal to
the experimental  value obtained  using QCD sum  rules.  To  date, the
aforementioned  modifications to  $V$  have been  carried  out at  the
two-loop  level only~\cite{Cornwall:1988ad}; clearly,  their all-order
generalization would be of great interest.

It is clear from the analysis presented that 
the role  of  the three-gluon vertex $\gv_{\mu\alpha\beta}$
vertex is  absolutely central. Specifically, the 
tensorial structure of the 
vertex, the presence or absence of kinematic poles,  
and the relative strength between the various components are 
determining factors for the existence of IR-finite solutions, 
and the type of running of the effective gluon mass.
As  we have mentioned in Sec.~\ref{Sect:IRF}, the Ansatz of (\ref{gtvertex})
employed for  the vertex attempts 
to  capture some  of the  main
features, but should be eventually obtained from an independent 
study of the dynamical SDE that it satisfies. 
The rich tensorial structure of a vertex with three Lorentz indices 
turns such a study into 
a rather complicated task, given that, in general, one has to deal   
with fourteen form-factors.
However, as a first approximation, one
can focus on those form-factors that enter into the expression determining 
$\Delta^{-1}(0)$, i.e. the generalized version of (\ref{Delta0}).
As an alternative, one could try to improve the  
Ans\"atze employed 
for the vertex, in the spirit of~\cite{King:1982mk},
in an attempt to correctly 
incorporate the required asymptotic behavior into the 
SDE, i.e. without having to resort to the heuristic procedure 
followed here. 

Another  source of  relative uncertainty  when evaluating  the  SDE in
question is the use of  the angular approximation.  As we explained in
detail, the  standard version of  this approximation gives rise  to an
approximate SDE that does not capture faithfully some of the essential
features  of the  original SDE.  To ameliorate  this drawback  we have
introduced  modifications to the  angular approximation,  presented in
the  Appendix.  In our opinion the  qualitative conclusions  of this
article are  robust and do  not depend on  its use; we expect  them to
persist  a  more complete  study,  where  the  angular integration  is
carried  out  numerically,  without  resorting to  any  approximation.
Should  quantitative discrepancies  arise, they  will  mainly manifest
themselves in  changes of  the values of  $c_1$ and $c_2$,  which, at
this  level of  approximation, are  free to  vary.  To  be  sure, once
dynamical information  for the  three-gluon vertex has  been furnished
one  would be  more  restricted, and  the two-dimensional  integration
should be carried out in its entirety.

Last but  not least,  let us turn  to the  possible role of  the ghost
sector.   As  we have  amply  emphasized,  in  the PT-BFM  scheme  the
omission   of  the   ghost   loops  does   not   interfere  with   the
gauge-invariance  of the  final  answer, and  in  particular with  the
transversality of the gluon self-energy. In addition, neglecting ghost
contributions only affects the  RG-logarithms by about only $10\%$, the
difference   between  $b=   11  C_A/48\pi^2$   and  ${\tilde   b}=  10
C_A/48\pi^2$. The possible impact of the ghosts in the IR is, however,
an  entirely different  matter; it  will greatly  depend,  among other
things, on  the structure and  solutions of the corresponding  SDE for
the ghost propagator within the PT-BFM formalism. Actually, given that
in this latter  framework one is working in the  Feynman gauge (of the
BFM) there is no a-priori reason that would exclude the possibility of
obtaining  IR-finite  solutions for  the  ghost  propagator.  If  this
turned  out  to be  true,  it would  suggest  that  in the  gluon-mass
description of QCD  the ghost sector may not play  such a central role
as   in   the   ``ghost-dominance''   picture~\cite{Atkinson:1997tu},
obtained  when   working  in  the  (conventional)   Landau  gauge.   A
preliminary study of these issues  is already underway, and we hope to
report its results in the near future.

\newpage

\appendix
\renewcommand{\theequation}{A-\arabic{equation}}
\section{\label{ang} Modified Angular Approximation}

In this Appendix we discuss the technical details related to 
the angular approximation and its modification.

It is convenient to introduce the following quantities, 
appearing on the r.h.s. of (\ref{sde2}):
\bea
I_0(q^2) &\equiv & q^2 \int d^4 k\,\Delta_0(k)\Delta(k+q)\,,
\nonumber\\
I_1(q^2) &\equiv & q^2 \int d^4 k\,\Delta(k) \Delta(k+q)\,,
\nonumber\\
I_2(q^2) &\equiv & \int d^4 k\, k^2 \Delta(k) \Delta(k+q)\,,
\nonumber\\
I_3(q^2) &\equiv & \frac{1}{q^2}
\int d^4 k\, k^2 \,\left[k^2 - (k+q)^2\right]\Delta(k) \Delta(k+q) \,.
\label{I0123}
\eea
Note that 
\be
\int d^4 k \frac{(k \cdot q)^2}{q^2} \Delta(k) \Delta(k+q) = 
\frac{1}{4}\, I_1(q^2) + \frac{1}{2}\, I_3(q^2)\,.
\ee
The angular integration of $I_0(q)$ may be carried out exactly,
after shifting the integration variable $k+q \to k$, 
\bea
I_0(q^2)  &=&  q^2 \int d^4 k\, \frac{\Delta(k)}{(k+q)^2}\nonumber\\
&=& \pi^2\bigg[ \int_{0}^{x} dy y \Delta(y)+ x \int_{x}^{\infty} dy \Delta(y)\bigg] \,.
\label{I0}
\eea
For the remaining integrals we will use the angular approximation,
appropriately modified to account correctly for their 
contributions to the running of the mass $m^2(x)$, 
i.e. terms that will enter in Eq.~(\ref{meq}).

The standard angular approximation amounts to 
\be
\int_{0}^{\pi}\!\!\! d\chi\sin^2\chi \, f(z) \approx \frac{\pi}{2} 
\bigg[\theta(x-y)f(x) + \theta(y-x)f(y)\bigg],
\label{angle}
\ee
with $z = x + y +2 \sqrt{xy}\cos\chi$, and $\theta(x)$ is the Heaviside step function.

Defining 
\be
{\bar I}_i(x)\equiv \pi^{-2} [I_i(x) - I_i(0)]\,, 
\ee
and using the superscript ``A''
to indicate that the aforementioned standard 
angular approximation has been employed, we obtain 
\bea
{\bar I}_1^{\chic A}(x) &=& x \Delta(x) \int_{0}^{x} \!\!\! dy\, y  \Delta(y)
+ x\int_{x}^{\infty} \!\!\!dy y \Delta^2(y) \,, 
\nonumber\\
{\bar I}_2^{\chic A}(x) &=& \Delta(x) \int_{0}^{x} \!\!\!  dy\, y^2  \Delta(y)
- \int_{0}^{x}\!\!\! dy y^2 \Delta^2(y)\,, 
\nonumber\\
{\bar I}_3^{\chic A}(x) &=& \frac{\Delta(x)}{x}\,\int_{0}^{x}\!\!\! dy\, y^2 (y-x)\Delta(y)\,.
\label{IiA}
\eea
Let us now check the faithfulness of these expressions for 
a special form of the massive propagator.
Specifically, substitute in (\ref{IiA}) $\Delta$ by the $\tilde\Delta$ of (\ref{tildeD}), 
apply the identity \mbox{$y\tilde\Delta(y) = 1 - m^2(y)\tilde\Delta(y)$}, 
keeping only the terms linear in $m^2(y)$
and set  $\tilde\Delta(x)\to x^{-1} $, to obtain 
\bea
{\bar I}_1^{\chic A}(x)|_{m^2} &=& - \int_{0}^{x}\!\!\!  dy \, m^2(y)  \tilde\Delta(y)
- x\int_{x}^{\infty}\!\!\! dy \, m^2(y) \tilde\Delta^2(y) \,,
\nonumber\\
{\bar I}_2^{\chic A}(x)|_{m^2} &=& 
2 \int_{0}^{x} \!\!\!dy \, m^2(y)  \tilde\Delta(y)
- \frac{1}{x}\int_{0}^{x} \!\!\! dy\, y \,  m^2(y)  \tilde\Delta(y) \,,
\nonumber\\
{\bar I}_3^{\chic A}(x)|_{m^2} &=& \frac{1}{x}\,\int_{0}^{x}\!\!\! dy \, y \, m^2(y)  \tilde\Delta(y)
- \frac{1}{x^2} \int_{0}^{x}\!\!\! dy \,y^2  m^2(y)  \tilde\Delta(y) \,.
\label{Iim}
\eea

Now the result in (\ref{Iim}) is to be compared with the direct calculation of 
 ${\bar I}_i(x)$, {\it before} 
resorting to (\ref{angle}). 
We begin with $I_2(q^2)$; by substituting $\Delta$ by $\tilde\Delta$ into 
$I_2(q^2)$ of (\ref{I0123}) and isolating again the mass terms, we have
\be
I_2(q^2) = \int d^4 k\,\tilde\Delta(k) - \int d^4 k\, m^2(k) \tilde\Delta(k) \tilde\Delta(k+q) \,,
\ee
and
\be
I_2(q^2) - I_2(0) = \int d^4 k\, m^2(k) \tilde\Delta(k)\left[\tilde\Delta(k)-\tilde\Delta(k+q)\right] \,.
\ee
This is the {\it exact mass dependence}. 
To this last expression  we now apply (\ref{angle}), to get
\bea
{\bar I}_2^{m^2}(x)|_{\chic A} &=& 
\int_{0}^{x}\!\!\! dy\, y \,m^2(y) \tilde\Delta(y)\left[\tilde\Delta(y) - \tilde\Delta(x)\right]
\nonumber\\
&=& \int_{0}^{x}\!\!\! dy \,m^2(y) \tilde\Delta(y) - \frac{1}{x}\int_{0}^{x} \!\!\! dy\, y \,  m^2(y)  \tilde\Delta(y)
\,\,+ {\cal O}(m^4) \,.
\label{i2m}
\eea
Comparing ${\bar I}_2^{m^2}(x)|_{\chic A}$ with  ${\bar I}_2^{\chic A}(x)|_{m^2}$ [Eqs.(\ref{i2m}) and (\ref{Iim})]  we  
see a discrepancy of a factor of $2$ in their first term.
The simplest term, {\it quadratic in $\Delta$}, that could 
correct this discrepancy is 
$\frac{1}{2}\int_{0}^{x} dy y^2 \Delta^2(y)$;
adding it to the result  of the standard angular approximation, ${\bar I}_2^{\chic A}(x)$ [Eq.(\ref{IiA})], 
will lead to a modified ${\bar I}_2^{\chic A}(x)|_{m^2}$, that will coincide 
with  ${\bar I}_2^{m^2}(x)|_{\chic A}$.  

We will next repeat the same exercise for $I_3(q^2)$. It is elementary to show, by substituting 
$\Delta \to \tilde\Delta$ into $I_3(q^2)$ that 
\be
I_3(q^2)
= \frac{1}{q^2}\int d^4 k\,\left[m^2(k+q)-m^2(k)\right] \,k^2
\tilde\Delta(k) \tilde\Delta(k+q) 
\,+ \,\int d^4 k\,\tilde\Delta(k) \,,
\ee
and therefore, applying the angular approximation and subtracting at $x=0$, 
\bea
{\bar I}_3^{m^2}(x)|_{\chic A} &=&  \frac{\tilde\Delta(x)}{x}
\int_{0}^{x}\!\!\! dy \,y^2 \left[m^2(x)-m^2(y)\right]\,\tilde\Delta(y) 
\nonumber\\
& \sim & \frac{m^2(x)}{2} - \frac{1}{x^2} \int_{0}^{x}\!\!\!dy\, y^2 m^2(y) \,\tilde\Delta(y)\,.
\eea
This term is subleading for either type of running of the mass 
(logarithmic or power-law). This important property is obviously 
{\it not} captured by the approximate expression of ${\bar I}_3^{\chic A}(x)|_{m^2}$ in
(\ref{Iim}), since the first term would furnish (erroneously) a leading order contribution 
in the case of power-law running. This shortcoming may be remedied
by simply adding to ${\bar I}_3^{\chic A}(x)$, Eq.(\ref{IiA}), the term $\frac{1}{2x}\int_{0}^{x} dy y^3 \Delta^2(y)$;
thus, the subleading nature of the term $I_3(q)$ is preserved.

We finally turn to $I_1(q^2)$.
The most immediate way to see that the standard angular  approximation for $I_1$, 
given by  the first equation in (\ref{IiA}),
mistreats the 
running of the mass, is to set $\Delta(y)\to \tilde\Delta(y)$ and then  $m^2(y)=m^2$
in the initial expression for $I_1(q^2)$ 
(second equation in (\ref{I0123})).
Then $I_1(q^2)$ gets reduced to a standard one-loop integral, 
\be
I_1(q^2) = q^2 \int\! \frac{d^4k}{(k^2+m^2)[(k+q)^2 + m^2]} \,,
\ee
and thus ${\bar I}_1(q^2)$ is given by
\be
{\bar I}_1(q^2)= - q^2 \Bigg[c-2 + \ln\left(\frac{m^2}{\mu^2}\right) + D \ln \left(\frac{D+1}{D-1}\right)\Bigg]\,,
\label{Iasy}
\ee
where $c= -\frac{2}{\epsilon} + \gamma -\ln 4\pi$ and 
$D=\left(1+\frac{4m^2}{q^2}\right)^{1/2}$.
For large values of $q^2$ we have that the finite part of ${\bar I}_1(q^2)$ goes like
\be
{\bar I}_1(x) \sim  - x\ln x  -  2 m^2\ln x \,.
\label{Iex}
\ee
Notice that the last term, which determines the contribution to the running of the 
mass, comes with weight 2 compared to the logarithmic term determining the 
asymptotic running of $\Delta^{\,-1}(x)$ (remember that 
$I_1(q^2)$  is to be multiplied by $(- \tilde b)$, so 
the first term of (\ref{Iasy}) provides the RG logarithm).
Instead, 
setting $\Delta(y)\to \tilde\Delta(y)$ and $m^2(y)=m^2$ in 
${\bar I}_1^{\chic A}(x)$ and  ${\bar I}_1^{\chic A}(x)|_{m^2}$, [Eqs.(\ref{IiA}) and (\ref{Iim})],
we find as leading contribution
\bea
{\bar I}_1^{\chic A}(x)|_{m^2} & \sim & - m^2 \ln x \nonumber\\
{\bar I}_1^{\chic A}(x) & \sim & -x\ln x - m^2 \ln x \,.
\eea
Unlike the (exact) expression of (\ref{Iex}), now the two logarithms
appear with equal relative weight.
Evidently, the angular approximation captures correctly 
the leading (RG) logarithm, but furnishes only half of the  $ m^2 \ln x$ contribution.

There is another way to verify that the correct 
relative weight between the two logarithms of $I_1(q^2)$ is 2. 
When the spectral representation for $\Delta (q)$ is used, 
\be
\Delta (q^2) = \int \!\! d \lambda^2 \, \frac{\rho\, (\lambda^2)}{q^2 - \lambda^2 + i\epsilon}\,,
\label{lehmann}
\ee
the ``kinetic'' term of the resulting (linearized) SDE, i.e. the exact analogue 
to $-I_1(q^2)$, is given by  
\be
B(q^2) = q^2 \int \!\! d \lambda^2 \, \rho\, (\lambda^2) \int\! d^4k \, \Delta (k)\,
\Delta (k+q)\,,
\ee
and may be easily cast in the form 
\be
B(x)=  c x + x \int^{x/4}_{0}\!\!\!dy  \left(1-\frac{4y}{x}\right)^{1/2} \Delta(y)\,, 
\ee
where $c$ is a (divergent) constant, to be absorbed into the 
wave-function renormalization of the gluon field.
When determining  the contribution of $B(x)$ to the running of $m^2$ 
for large $x$, it would be
wrong to simply set $\left(1-\frac{4y}{x}\right)^{1/2} \to 1$; instead,
one must expand this term to first order. Setting $\Delta(y)\to \tilde\Delta(y)$ 
we obtain
\bea
B(x) &=&  c x + x \int^{x/4}_{0}\!\!\!dy \tilde\Delta(y)\ 
- 2\int^{x/4}_{0}\!\!\!dy \,y \tilde\Delta(y)\nonumber\\
 &\sim & c^{\prime} x  + x \ln x + 2 \int^{x/4}_{0}\!\!\!dy\, m^2(y)\tilde\Delta(y)\,,
\eea
where $c^{\prime}$ includes now some irrelevant (finite) constant term.
Again, the mass term is multiplied by a factor of 2 compared to the RG logarithm.

The simplest way of remedy this discrepancy is to modify the result of the 
angular approximation for $I_1(q^2)$, adding  the term 
$\frac{1}{2}\int_{0}^{x} dy y^2 \Delta^2(y)$; note that 
this is exactly the term that 
has been added to $I_2(q^2)$.

Thus, finally, we arrive at the following modified expressions, which 
capture correctly the leading $m^2$-dependence of the SDE: 
\bea
{\bar I}_1^{\chic{MA}}(x) &=&  x\Delta(x) \int_{0}^{x} \!\!\! dy\, y  \Delta(y)
+\frac{1}{2}\int_{0}^{x} dy y^2 \Delta^2(y) + x\int_{x}^{\infty}\!\!\! dy\, y \Delta^2(y)\,, 
\nonumber\\
{\bar I}_2^{\chic{MA}}(x)   &=& \Delta(x) \int_{0}^{x} \!\!\! dy \,y^2  \Delta(y)
- \frac{1}{2}\int_{0}^{x}\!\!\! dy\, y^2 \Delta^2(y)\,,
\nonumber\\
{\bar I}_3^{\chic{MA}}(x) &=& 
 \frac{\Delta(x)}{x} \int_{0}^{x} \!\!\!dy \,y^2 (y-x)\Delta(y)
+ \frac{1}{2x}\int_{0}^{x}\!\!\! dy\, y^3 \Delta^2(y)\,.
\label{mang}
\eea

\section*{Acknowledgments}
This work was supported by the Spanish MEC under the grants FPA 2005-01678 and 
FPA 2005-00711.  
The research of JP is funded by the Fundaci\'on General of the UV. 
The authors thank Professor J~.M~.Cornwall for several useful communications.


\begin{thebibliography}{99}

\bibitem{Cornwall:1979hz}
  J.~M.~Cornwall,
  Nucl.\ Phys.\ B {\bf 157}, 392 (1979).

\bibitem{Cornwall:1981zr}
  J.~M.~Cornwall,
  Phys.\ Rev.\  D {\bf 26}, 1453 (1982).


\bibitem{Bernard:1982my}
  C.~W.~Bernard,
  Nucl.\ Phys.\ B {\bf 219}, 341 (1983);
  J.~F.~Donoghue,
  Phys.\ Rev.\ D {\bf 29}, 2559 (1984);
  M.~H.~Thoma and H.~J.~Mang,
  Z.\ Phys.\ C {\bf 44}, 349 (1989);
  E.~Bagan and M.~R.~Pennington,
  Phys.\ Lett.\ B {\bf 220}, 453 (1989);
  U.~Habel, R.~Konning, H.~G.~Reusch, M.~Stingl and S.~Wigard,
  Z.\ Phys.\ A {\bf 336}, 423 (1990);
  Z.\ Phys.\ A {\bf 336} (1990) 435;
  J.~E.~Shrauner,
  J.\ Phys.\ G {\bf 19}, 979 (1993).


\bibitem{Kondo:2001nq}
  K.~I.~Kondo,
  Phys.\ Lett.\ B {\bf 514}, 335 (2001);
  K.~I.~Kondo, T.~Murakami, T.~Shinohara and T.~Imai,
  Phys.\ Rev.\ D {\bf 65}, 085034 (2002); 
  K.~I.~Kondo,
  Phys.\ Rev.\  D {\bf 74}, 125003 (2006)


  
\bibitem{Bloch:2003yu}
  J.~C.~R.~Bloch,
  Few Body Syst.\  {\bf 33}, 111 (2003).

\bibitem{Aguilar:2004sw}
  A.~C.~Aguilar and A.~A.~Natale,
  JHEP {\bf 0408}, 057 (2004).




\bibitem{Dudal:2004rx}
  D.~Dudal, J.~A.~Gracey, V.~E.~R.~Lemes, M.~S.~Sarandy, R.~F.~Sobreiro, S.~P.~Sorella and H.~Verschelde,
  Phys.\ Rev.\ D {\bf 70}, 114038 (2004);
  D.~Dudal, H.~Verschelde, J.~A.~Gracey, V.~E.~R.~Lemes, M.~S.~Sarandy, R.~F.~Sobreiro and S.~P.~Sorella,
  JHEP {\bf 0401}, 044 (2004);
  S.~P.~Sorella,
  Annals Phys.\  {\bf 321}, 1747 (2006).

\bibitem{Aguilar:2006gr}
  A.~C.~Aguilar and J.~Papavassiliou,
  JHEP {\bf 0612}, 012 (2006).





\bibitem{Parisi:1980jy}
  G.~Parisi and R.~Petronzio,
  Phys.\ Lett.\  B {\bf 94}, 51 (1980).

\bibitem{Mattingly:1992ud}
  A.~C.~Mattingly and P.~M.~Stevenson,
  Phys.\ Rev.\ Lett.\  {\bf 69}, 1320 (1992);
  F.~Halzen, G.~I.~Krein and A.~A.~Natale,
  Phys.\ Rev.\ D {\bf 47}, 295 (1993);
  M.~B.~Gay Ducati, F.~Halzen and A.~A.~Natale,
  Phys.\ Rev.\ D {\bf 48}, 2324 (1993);
  J.~R.~Cudell and B.~U.~Nguyen,
  Nucl.\ Phys.\ B {\bf 420}, 669 (1994);
  M.~Consoli and J.~H.~Field,
  Phys.\ Rev.\ D {\bf 49}, 1293 (1994);
  F.~J.~Yndurain,
  Phys.\ Lett.\ B {\bf 345}, 524 (1995);
  A.~Szczepaniak, E.~S.~Swanson, C.~R.~Ji and S.~R.~Cotanch,
  Phys.\ Rev.\ Lett.\  {\bf 76}, 2011 (1996).
  A.~Donnachie and P.~V.~Landshoff,
  Phys.\ Lett.\ B {\bf 387}, 637 (1996);
  M.~Anselmino and F.~Murgia,
  Phys.\ Rev.\  D {\bf 53}, 5314 (1996);
  M.~Consoli and J.~H.~Field,
  J.\ Phys.\ G {\bf 23}, 41 (1997).



\bibitem{Mihara:2000wf}
  A.~Mihara and A.~A.~Natale,
  Phys.\ Lett.\ B {\bf 482}, 378 (2000);
J.~H.~Field,
  Phys.\ Rev.\ D {\bf 66}, 013013 (2002);
  F.~Cano and J.~M.~Laget,
  Phys.\ Rev.\ D {\bf 65}, 074022 (2002);
  R.~Enberg, G.~Ingelman and L.~Motyka,
  Phys.\ Lett.\ B {\bf 524}, 273 (2002);
  M.~B.~Gay Ducati and W.~K.~Sauter,
  Phys.\ Rev.\ D {\bf 67}, 014014 (2003);
  W.~S.~Hou and G.~G.~Wong,
  Phys.\ Rev.\ D {\bf 67}, 034003 (2003);
  E.~G.~S.~Luna, A.~F.~Martini, M.~J.~Menon, A.~Mihara and A.~A.~Natale,
  Phys.\ Rev.\ D {\bf 72}, 034019 (2005);
  E.~G.~S.~Luna and A.~A.~Natale,
  Phys.\ Rev.\ D {\bf 73}, 074019 (2006);
  E.~G.~S.~Luna, A.~A.~Natale and C.~M.~Zanetti,
  arXiv:hep-ph/0605338;
  E.~G.~S.~Luna,
  Phys.\ Lett.\  B {\bf 641}, 171 (2006).


\bibitem{lattice}
In  addition, the non-perturbative behavior of QCD Green's functions
found in 
lattice simulations may be described 
in terms of effectively massive gluon  propagators,
see, for example,
  C.~Alexandrou, P.~de Forcrand and E.~Follana,
  Phys.\ Rev.\ D {\bf 63}, 094504 (2001);
  Phys.\ Rev.\ D {\bf 65}, 117502 (2002);
  Phys.\ Rev.\ D {\bf 65}, 114508 (2002);
  F.~D.~R.~Bonnet, P.~O.~Bowman, D.~B.~Leinweber and A.~G.~Williams,
  Phys.\ Rev.\ D {\bf 62}, 051501 (2000);
  F.~D.~R.~Bonnet, P.~O.~Bowman, D.~B.~Leinweber, A.~G.~Williams and J.~M.~Zanotti,
  Phys.\ Rev.\ D {\bf 64}, 034501 (2001);
  A.~Sternbeck, E.~M.~Ilgenfritz, M.~Mueller-Preussker and A.~Schiller,
  Phys.\ Rev.\ D {\bf 72}, 014507 (2005);
  P.~J.~Silva and O.~Oliveira,
  Phys.\ Rev.\  D {\bf 74}, 034513 (2006);
  Ph.~Boucaud {\it et al.},
  arXiv:hep-ph/0507104;
  Ph.~Boucaud {\it et al.},
  JHEP {\bf 0606}, 001 (2006);
  A.~Cucchieri and T.~Mendes,
  Phys.\ Rev.\ D {\bf 73}, 071502 (2006).

\bibitem{Schwinger:1962tn}
  J.~S.~Schwinger,
  Phys.\ Rev.\  {\bf 125}, 397 (1962);
  Phys.\ Rev.\  {\bf 128}, 2425 (1962).




\bibitem{Shifman:1978by}
  M.~A.~Shifman, A.~I.~Vainshtein and V.~I.~Zakharov,
  Nucl.\ Phys.\ B {\bf 147}, 448 (1979);
  Nucl.\ Phys.\ B {\bf 147}, 385 (1979).


\bibitem{Cornwall:1989gv}
J.~M.~Cornwall and J.~Papavassiliou,
Phys.\ Rev.\ D {\bf 40}, 3474 (1989).


\bibitem{Binosi:2002ft}
  D.~Binosi and J.~Papavassiliou,
  Phys.\ Rev.\ D {\bf 66}, 111901 (2002);
  J.\ Phys.\ G {\bf 30}, 203 (2004).

\bibitem{Abbott:1980hw}
L.~F.~Abbott,
Nucl.\ Phys.\ B {\bf 185}, 189 (1981).

\bibitem{Binosi:2006da}
  D.~Binosi and J.~Papavassiliou,
  JHEP {\bf 0703}, 041 (2007).



\bibitem{Papavassiliou:1991hx}
  J.~Papavassiliou and J.~M.~Cornwall,
  Phys.\ Rev.\ D {\bf 44}, 1285 (1991).
 

\bibitem{Badalian:1999fq}
  A.~M.~Badalian and V.~L.~Morgunov,
  Phys.\ Rev.\ D {\bf 60}, 116008 (1999).

\bibitem{Aguilar:2002tc}
  A.~C.~Aguilar, A.~A.~Natale and P.~S.~Rodrigues da Silva,
  Phys.\ Rev.\ Lett.\  {\bf 90}, 152001 (2003);
  A.~C.~Aguilar, A.~Mihara and A.~A.~Natale,
  Phys.\ Rev.\ D {\bf 65}, 054011 (2002);
  Int.\ J.\ Mod.\ Phys.\ A {\bf 19} (2004) 249.



\bibitem{Brodsky:2002nb}
  S.~J.~Brodsky, S.~Menke, C.~Merino and J.~Rathsman,
  Phys.\ Rev.\ D {\bf 67}, 055008 (2003);
  S.~J.~Brodsky,
  Fizika B {\bf 13}, 91 (2004).
  
  



\bibitem{Mattingly:1993ej}
The freezing of the QCD coupling has also been 
advocated in various different approaches, e.g.,
A.~C.~Mattingly and P.~M.~Stevenson,
  Phys.\ Rev.\ D {\bf 49}, 437 (1994);
Y.~L.~Dokshitzer, G.~Marchesini and B.~R.~Webber,
Nucl.\ Phys.\ B {\bf 469}, 93 (1996); 
  L.~von Smekal, R.~Alkofer and A.~Hauck,
  Phys.\ Rev.\ Lett.\  {\bf 79}, 3591 (1997);
  M.~Baldicchi and G.~M.~Prosperi,
  Phys.\ Rev.\ D {\bf 66}, 074008 (2002);
  G.~Grunberg,
  Phys.\ Rev.\ D {\bf 29}, 2315 (1984);
  Phys.\ Rev.\ D {\bf 73}, 091901 (2006);
  H.~Gies,
  Phys.\ Rev.\ D {\bf 66}, 025006 (2002);
  D.~V.~Shirkov and I.~L.~Solovtsov,
  Phys.\ Rev.\ Lett.\  {\bf 79}, 1209 (1997);
  A.~V.~Nesterenko and J.~Papavassiliou,
  Phys.\ Rev.\ D {\bf 71}, 016009 (2005);
  A.~P.~Bakulev, S.~V.~Mikhailov and N.~G.~Stefanis,
  Phys.\ Rev.\ D {\bf 72}, 074014 (2005);
  J.~A.~Gracey,
  JHEP {\bf 0605}, 052 (2006);
  G.~M.~Prosperi, M.~Raciti and C.~Simolo,
  Prog.\ Part.\ Nucl.\ Phys.\  {\bf 58}, 387 (2007).



\bibitem{Lane:1974he}
See, for example,  K.~D.~Lane,
  Phys.\ Rev.\ D {\bf 10}, 2605 (1974);
  H.~Pagels,
  Phys.\ Rev.\ D {\bf 19}, 3080 (1979);
  V.~A.~Miransky,
  Phys.\ Lett.\  B {\bf 165}, 401 (1985);
  C.~D.~Roberts and A.~G.~Williams,
  Prog.\ Part.\ Nucl.\ Phys.\  {\bf 33}, 477 (1994).


\bibitem{Cornwall:1985bg}
  J.~M.~Cornwall and W.~S.~Hou,
  Phys.\ Rev.\  D {\bf 34}, 585 (1986).

\bibitem{Lavelle:1991ve}
  M.~Lavelle,
  Phys.\ Rev.\  D {\bf 44}, 26 (1991).


\bibitem{com3} It  is important to notice that  the conventional gluon
self-energy contains in addition
unphysical  condensates involving  ghost operators, see,  
  M.~J.~Lavelle and M.~Schaden,
  Phys.\ Lett.\  B {\bf 208}, 297 (1988);
  E.~Bagan and T.~G.~Steele,
  Phys.\ Lett.\  B {\bf 219}, 497 (1989).
Such condensates
cancel out exactly against the propagator-like contributions contained
in vertices and boxes, extracted following the standard PT 
procedure~\cite{Lavelle:1991ve}.

\bibitem{Sohn:1985em}
The full 
SDE for the BFM gluon self-energy was first derived 
in~
R.~B.~Sohn,
  Nucl.\ Phys.\ B {\bf 273}, 468 (1986);
  A.~Hadicke,
JENA-N-88-19.

\bibitem{Binosi:2002ez}
D.~Binosi and J.~Papavassiliou,
Phys.\ Rev.\ D {\bf 66}, 025024 (2002)


\bibitem{Gambino:1999ai}
  P.~Gambino and P.~A.~Grassi,
  Phys.\ Rev.\ D {\bf 62}, 076002 (2000);
  P.~A.~Grassi, T.~Hurth and M.~Steinhauser,
  Annals Phys.\  {\bf 288}, 197 (2001).

\bibitem{foot1}
Note in passing that this type of   
generalized Feynman gauge cannot be obtained through an appropriate choice of the 
(constant) gauge-fixing parameter $\xi$. Instead, it is reminiscent of the  
so-called ``stagnant gauge'', presented in 
  C.~H.~Llewellyn Smith,
  Nucl.\ Phys.\  B {\bf 165}, 423 (1980);
it may be formally reached by introducing in the 
Feynman diagrams a momentum dependent  $\xi(q^2)$,
or an operator $\xi(\Box)$ in the Lagrangian.  

\bibitem{Salam:1963sa}
A.~Salam,
Phys.\ Rev.\  {\bf 130}, 1287 (1963); 
R.~Delbourgo and A.~Salam,
Phys.\ Rev.\  {\bf 135}, B1398 (1964);     
R.~Delbourgo and P.~West,
J.\ Phys.\ A {\bf 10}, 1049 (1977); 
R.~Delbourgo,
Nuovo Cim.\ A {\bf 49}, 484 (1979). 



\bibitem{Jackiw:1973tr}
  R.~Jackiw and K.~Johnson,
  Phys.\ Rev.\ D {\bf 8}, 2386 (1973);
  J.~M.~Cornwall and R.~E.~Norton,
  Phys.\ Rev.\ D {\bf 8}, 3338 (1973);
  E.~Eichten and F.~Feinberg,
  Phys.\ Rev.\ D {\bf 10}, 3254 (1974).


\bibitem{Ball:1980ax}
  J.~S.~Ball and T.~W.~Chiu,
  Phys.\ Rev.\ D {\bf 22}, 2550 (1980),
  [Erratum-ibid.\ D {\bf 23}, 3085 (1981)].

\bibitem{Binger:2006sj}
  M.~Binger and S.~J.~Brodsky,
  Phys.\ Rev.\ D {\bf 74}, 054016 (2006).


\bibitem{Binosi:2002vk}
  D.~Binosi and J.~Papavassiliou,
  Nucl.\ Phys.\ Proc.\ Suppl.\  {\bf 121}, 281 (2003)


\bibitem{King:1982mk}
J.~E.~King,
  Phys.\ Rev.\ D {\bf 27}, 1821 (1983);
  B.~J.~Haeri,
  Phys.\ Rev.\ D {\bf 38}, 3799 (1988).


\bibitem{com2} 
The mass scale $\mu_2$ is associated with the quark condensate 
 $\langle{\bar\psi}\psi \rangle$ of dimension three, while  
$\mu_1$ with $M_0$, a bare quark mass that breaks 
chiral symmetry explicitly.

\bibitem{GR}
See, for example, 
I.S.~Gradshteyn and I.M.~Ryzhik, ``Table of Integrals, Series, and 
and Products'', Fifth Edition, Academic Press, London.


\bibitem{Graziani:1984cs}
  F.~R.~Graziani,
  Z.\ Phys.\  C {\bf 33}, 397 (1987). 



\bibitem{Kogan:1994wf}
  I.~I.~Kogan and A.~Kovner,
  Phys.\ Rev.\  D {\bf 52}, 3719 (1995).

\bibitem{Gorbar:1999xi}
  E.~V.~Gorbar and A.~A.~Natale,
  Phys.\ Rev.\  D {\bf 61}, 054012 (2000).

\bibitem{Cornwall:1974vz}
  J.~M.~Cornwall, R.~Jackiw and E.~Tomboulis,
  Phys.\ Rev.\ D {\bf 10}, 2428 (1974).



\bibitem{Cornwall:1988ad}
  J.~M.~Cornwall,
  Physica A {\bf 158}, 97 (1989).




\bibitem{amin}
In addition to  $\langle  G^2 \rangle$, 
another quantity that may be relevant 
to these considerations is 
the {\it gauge-invariant non-local} condensate of dimension two,
usually denoted by $\langle A^2 _{\rm min}\rangle$, obtained through the minimization
of $\int d^4 x ( A_{\mu})^2$ over all gauge transformations
\cite{Gubarev:2000eu,Gracey:2007ki}, or variants of it 
involving also ghost condensates~\cite{Kondo:2001nq}.
$\langle A^2 _{\rm min}\rangle$ should not to be confused 
with $\langle 0|\!:\!A_{\mu}^{a} A^{\mu}_{a}\!:\!|0 \rangle$, the local  
{\it gauge-variant} condensate of dimension two; 
the latter cannot appear in the OPE of gauge-invariant quantities. 

\bibitem{Gubarev:2000eu}
  F.~V.~Gubarev, L.~Stodolsky and V.~I.~Zakharov,
  Phys.\ Rev.\ Lett.\  {\bf 86}, 2220 (2001);
  F.~V.~Gubarev and V.~I.~Zakharov,
  Phys.\ Lett.\  B {\bf 501}, 28 (2001).

\bibitem{Gracey:2007ki}
  J.~A.~Gracey,
  arXiv:0706.1440 [hep-th], and references therein.





\bibitem{Atkinson:1997tu}
  D.~Atkinson and J.~C.~R.~Bloch,
Phys.\ Rev.\ D {\bf 58}, 094036 (1998);
  R.~Alkofer and L.~von Smekal,
  Phys.\ Rept.\  {\bf 353}, 281 (2001);
  J.~C.~R.~Bloch,
  Phys.\ Rev.\ D {\bf 64}, 116011 (2001);
  R.~Alkofer, C.~S.~Fischer and F.~J.~Llanes-Estrada,
  Phys.\ Lett.\ B {\bf 611}, 279 (2005);
  C.~S.~Fischer,
  J.\ Phys.\ G {\bf 32}, R253 (2006).
 


\end{thebibliography}
\end{document}